\documentclass[manuscript]{acmart}

\AtBeginDocument{%
  \providecommand\BibTeX{{%
    \normalfont B\kern-0.5em{\scshape i\kern-0.25em b}\kern-0.8em\TeX}}}
\newcommand{\inlinequote}[2]{\emph{``#1''} (\textit{I#2})}

\newcommand\citetodo[1]{\textcolor{olive}{[CITE]}}

\usepackage{multirow}

\usepackage[utf8]{inputenc}
\usepackage{enumitem}
\usepackage{csquotes}
\usepackage{wrapfig}

\usepackage[para]{footmisc}
\usepackage{makecell}

\usepackage{tcolorbox}
\usepackage{lipsum}

\usepackage[false]{anonymous-acm}

\usepackage{graphicx}
\usepackage{subcaption}
\usepackage{float}
\usepackage[export]{adjustbox}

\definecolor{HighEfficacy}{HTML}{B8E7DE}
\definecolor{AssistedSuccess}{HTML}{D9E7E8}
\definecolor{UncertainPerformance}{HTML}{FDF0D6}
\definecolor{Underperforming}{HTML}{F7D1AE}
\definecolor{NegativeUtility}{HTML}{F0C2BE}
\definecolor{QuestionableIntegrity}{HTML}{E0D5F6}
\definecolor{NotMentioned}{HTML}{f3f6f4}

\definecolor{todocolor}{HTML}{EF2D56}

\usepackage[normalem]{ulem}
\useunder{\uline}{\ul}{}

\copyrightyear{2025}
\acmYear{2025}
\setcopyright{rightsretained}
\acmConference[DIS '25]{DIS '25}{July 5-9,
2025}{Madeira, Portugal}
\acmDOI{}
\acmISBN{}
\acmSubmissionID{4581}

\begin{document}

\title[Co-Designing with Algorithms]{Co-Designing with Algorithms: Unpacking the Complex Role of GenAI in Interactive System Design Education}

\author{Hauke Sandhaus}
\email{hgs52@cornell.edu}
\orcid{0000-0002-4169-0197}
\affiliation{%
    \institution{Cornell University, Cornell Tech}
    \streetaddress{2 West Loop Rd}
    \city{New York}
    \state{New York}
    \country{USA}
    \postcode{10044}
}
\author{Quiquan Gu}
\orcid{0009-0000-9544-7654}
\email{guqq@berkeley.edu}
\affiliation{%
  \institution{UC Berkeley}
  \streetaddress{200 California Hall}
  \city{Berkeley}
  \state{California}
  \country{USA}
  \postcode{94720}
}

\author{Maria Teresa Parreira}
\orcid{}
\email{mb2554@cornell.edu}
\affiliation{%
    \institution{Cornell University, Cornell Tech}
    \streetaddress{2 West Loop Rd}
    \city{New York}
    \state{New York}
    \country{USA}
    \postcode{10044}
}
\author{Wendy Ju}
\orcid{0000-0002-3119-611X}
\email{wendyju@cornell.edu}
\affiliation{%
  \institution{Cornell Tech}
  \streetaddress{2 West Loop Rd}
  \city{New York}
  \state{New York}
  \country{USA}
  \postcode{10044}
}

\renewcommand{\shortauthors}{Sandhaus, Gu, Parreira, and Ju}

\begin{abstract}
Generative Artificial Intelligence (GenAI) is transforming Human-Computer Interaction (HCI) education and technology design, yet its impact remains poorly understood. This study explores how graduate students in an applied HCI course used GenAI tools during interactive device design. Despite no encouragement, all groups integrated GenAI into their workflows. Through 12 post-class group interviews, we identified how GenAI co-design behaviors present both benefits—such as enhanced creativity and faster design iterations—and risks, including shallow learning and reflection. Benefits were most evident during the execution phases, while the discovery and reflection phases showed limited gains. A taxonomy of usage patterns revealed that students' outcomes depended more on \emph{how} they used GenAI than the specific tasks performed. These findings highlight the need for HCI education to adapt to GenAI’s role and offer recommendations for curricula to better prepare future designers for effective creative co-design.

\end{abstract}



\begin{teaserfigure}
    \centering
    \includegraphics[width=0.95\linewidth]{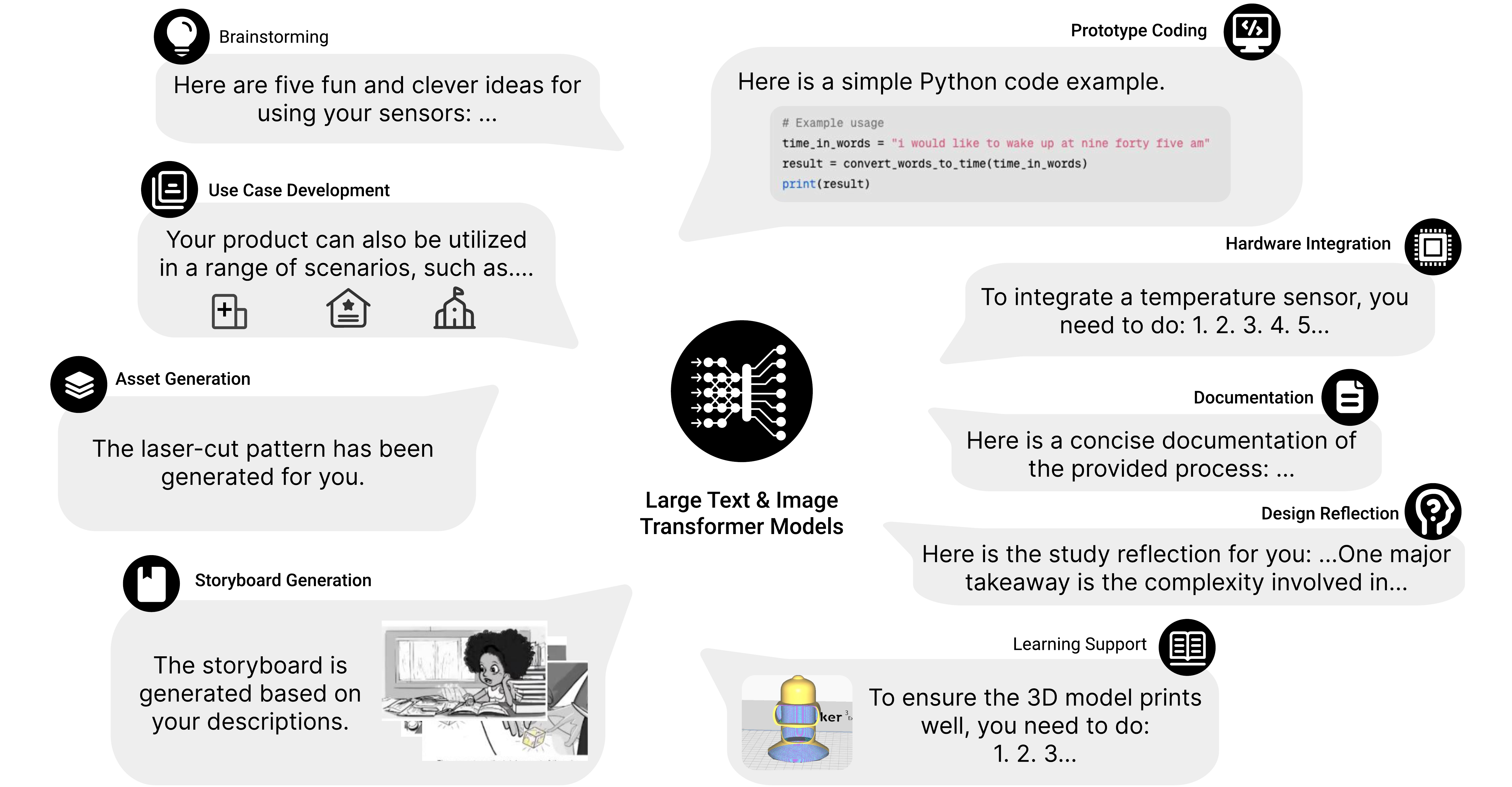}
    \caption{Identified most common use cases of GenAI in the Interactive Device Design class, exemplified by chatbot-style responses.}
    \label{fig:use-case-examples}
    \Description{This figure showcases various use cases of GenAI in an HCI class, as depicted by chatbot-style responses. The central figure is a GenAI icon (transformer model) surrounded by dialogue bubbles representing different tasks. These tasks include brainstorming with a suggestion for five clever ideas for sensor use, prototype coding assistance with a simple Python code example, integrating a temperature sensor, asset generation with laser-cut pattern creation, storyboard generation based on descriptions, use case development for product scenarios, documentation assistance, design reflection, and learning support for 3D printing. Each task is associated with a different area of design work, with small icons representing the specific task, such as a lightbulb for brainstorming and a document for documentation assistance.}
\end{teaserfigure}

\received{20 January 2025}
\received[revised]{22 April 2025}

\maketitle

\section{Introduction}

The sudden emergence of Generative AI has educators scrambling to adapt; it is not clear how and when these tools are helpful or harmful. As these powerful tools become increasingly accessible to students, there is an urgent need to study GenAI's impact on the educational landscape. %
Can these technologies be incorporated into educational environments without compromising fundamental skill development? To ensure positive educational outcomes, it is crucial to critically examine and adapt pedagogical approaches to harness the benefits of GenAI while addressing potential challenges such as over-reliance or ethical concerns.

To better understand this phenomenon, we interviewed student groups about their use of GenAI tools for coursework. At the outset of the Fall 2023 graduate-level university course on interactive device design, students were instructed that they were permitted but not encouraged to use novel GenAI tools like GPT-4 and DALL-E in their coursework, provided that they honestly document their usage of these tools, just as they would document the use of instructions from Youtube or software libraries found on GitHub. This class consisted of six device design labs and a final project\footnote{Class Repository: \linkanon{https://github.com/FAR-Lab/Interactive-Lab-Hub/tree/Fall2023}{https://github.com/FAR-Lab/Interactive-Lab-Hub}}; students spent their time almost solely on project work by defining and executing design ideas for interactive devices. Following the semester (and, importantly, after grades were submitted and posted) students were interviewed about their self-initiated use of GenAI tools. While the selection of the class on interactive device design for HCI was a convenience sample, authors have taught similar courses at other institutions, and the course material features a good mix of ideation and technical prototyping skills, which presents a diversity of tasks that the GenAI tools might be employed for. This can help us understand how such tools might be used for good or ill in the context of prototyping classes in particular and HCI education in general. %

In this paper, we present our analysis and findings from these interviews, focusing on three research questions: \emph{(1) How do students use GenAI in HCI design?, (2) What are students' perceptions of GenAI usage?,} and \emph{(3) What are the potential long-term impacts on HCI education?}. We found that students engaged with GenAI throughout the entire design process. 
Students reported it to be most useful in the execution phase of the design process; it helped students by enabling faster iterations and diversifying the perspectives the students considered. %
In contrast, the use of GenAI tools was less fruitful in the definition stage of the design process. 

As part of this work, we contribute with a categorization of the main use cases of GenAI tools in HCI design, and map them onto the Double Diamond Design process \cite{designcouncilDoubleDiamondDesign2004}. We define a taxonomy of usage patterns of GenAI tools based on the role these tools take in relation to the student's skills and learning goals: Benchmark, Booster, Executer, and Amplifier.  This taxonomy proved helpful as we explored students' sentiments on the use of GenAI across the use cases and defined categories based on perceived efficiency and ethical use, revealing factors that impact these sentiments. 

Finally, we discuss GenAIs impact on HCI education, including analyses of students' performance, adherence to class policy, and self-reported reflections on the contribution of these tools to their design output.
While student's sentiment towards GenAI was favorable, students reflected that GenAI did not help them think critically about their approach and choices. This suggests a risk of diminished learning depth and reduced engagement in reflective thinking. This offers insights into how GenAI can both empower and handicap the pedagogical goals of HCI education, a field under the ongoing tension of teaching useful professional skills and values.

\section{Related Work}

Although the GenAI technology in question has only very recently become widely available and popular, there is already a growing body of work studying the impact this work has on HCI Design.

\subsection{GenAI’s Promise for Technology Design and HCI Practice}

Generative Artificial Intelligence (GenAI) holds significant potential for improving various stages of technology design and HCI practices. Recent workshops and studies have outlined the possibilities of using GenAI to facilitate creative human-technology collaboration~\cite{muller_genaichi_2023, shin_integrating_2023}. For example, GenAI's role in co-creation allows designers to ideate faster and iterate more efficiently~\cite{kulkarniWordWorthThousand2023}. HCI designers are exploring the ways GenAI can be used for speculative and critical design purposes, such as Design Fiction~\cite{blythe_artificial_2023, huang_future_2023}. This technology has also been praised for enhancing interaction design practices, where AI tools can help propose novel interactions that designers might not immediately consider~\cite{shi_hci-centric_2024}. Researchers speculate how GenAI might support designers as a long-term collaborator~\cite{vaithilingam_imagining_2024}, and be included in HCI design tasks~\cite{tholander_design_2023}. In addition to creative support, AI has shown its ability to assist in time-intensive tasks such as qualitative analysis, enabling researchers to conduct data-driven work more efficiently~\cite{xiao_supporting_2023, yan_human-ai_2024, yan_human-ai_2024}. 

Moreover, in applied technology design contexts, GenAI tools have already demonstrated their ability to accelerate processes like software coding, thereby enabling quicker prototyping~\cite{schmidt_speeding_2023}. 
Practitioners are beginning to integrate GenAI into the design process, as shown in a study by \citet{li_user_2024}.The study identifies eight key ways GenAI supports design tasks: (1) summarizing and prioritizing business requirements, (2) providing best practices for user requirements, (3) aiding in ideation with UI inspirations, (4) generating wireframes and automating responsive design, (5) conducting usability assessments during user testing, (6) suggesting design improvements, (7) creating presentation materials, and (8) organizing assets for development teams during design handoff. %

GenAI tools promise  HCI professionals to manage the growing complexity of technological systems and shorten innovation cycles, promising a more efficient workflow in design-driven practices. While some of these surprising uses of GenAI, a tool initially created for text generation, may be attributed to AI hype, it appears that many of its abilities are long lasting and cannot be only explained by early enthusiasm and a novelty effect~\cite{long_not_2024}. As GenAI appears to have long lasting effect on workflows, designers need to find beneficial ways to appropriate it~\cite{dixDesigningAppropriation2007}.

\subsection{GenAI’s Risks for Technology Design and Education}
Although GenAI promises to improve efficiency and creativity, its integration into HCI design also presents significant risks. While the release of GenAI ChatGPT version 4.0 was only recent, its rapid introduction poses societal risks that are presumably far reaching~\cite{baldassarre_social_2023}. There is still little research on how GenAI will reshape HCI technology design practice, but so far researchers found the impact of autonomous AI systems on data science practice to be inevitable, multi-layered, and call for new ways of human AI collaboration~\cite{wang_human-ai_2019}. Specifically, \citet{takaffoli_generative_2024} interviewed UX practitioners about their use of GenAI in the industry. They discovered a lack of policies, practices, and discussions regarding GenAI usage at both the company and team levels. Instead, individual UX practitioners often made independent decisions to use GenAI tools in their work and were also vigilant about data privacy and confidentiality. An interview study conducted by \citet{li_user_2024} indicates that junior designers may face negative effects such as skill degradation, job displacement, and a decline in creativity.

Other works have looked into how to design GenAI technology through human and user-centered design processes~\cite{weisz_design_2024} and identified numerous challenges. Output uncertainty~\cite{yang_re-examining_2020}, ethical concerns~\cite{xuAIDrivenUXUI2024}, biases embedded within AI systems \cite{li2024sustainable}, and the impact on future employment are critical risks that researchers and practitioners must contend with \cite{li_user_2024}. Additionally, research has shown that non-AI experts often struggle to effectively prompt AI systems~\cite{Zamfirescu-Pereira2023-sj}.

Education has been slow to address the challenges posed by GenAI, with student perspectives underrepresented in the discourse surrounding it~\cite{sullivan_chatgpt_2023}. As GenAI is capable to create passing grades for undergraduate coding assignments~\cite{richards_bob_2023}, some schools’ initial responses have been to \emph{ban its use}, while also increasing the weight of exam scores relative to take-home assignments~\cite{lau_ban_2023}. \citeauthor{lau_ban_2023} finds that programming instructors generally lean toward two long-term plans: either embracing AI coding tools by integrating them into their courses or resisting their use altogether. Instructors advocating for resistance stress the significance of coding fundamentals, ethical issues, and AI-proof assessments by switching to oral, paper-based, or image-based exams.

Looking beyond GenAI for coding education, GenAI is mostly seen through the concern for plagiarism and cheating~\cite{cotton_chatting_2024}. A literature review on GenAI's impact on education highlights issues of accuracy, reliability, and plagiarism, noting that institutions are trying to identify GenAI content, create policies, and implement GenAI-proof tasks like digital-free assessments and novel question types~\cite{lo_what_2023}. To sum up, class policies responding to GenAI can be broadly grouped into three categories: ban the technology, integrate and promote it, or allow it in specific contexts and with acknowledgement~\cite{carnegiemellonuniversityExamplesPossibleAcademic2024, emilycallahanExamplesClassroomPolicies2024, centerforteaching&learningtheuniversityoftexasataustinChatGPTGenerativeAI2024}.

\subsection{Positive Impact of GenAI on Novice HCI Practice}
While there has been research on the use of GenAI in professional design practice, less is known about how novice HCI practitioners (such as students) engage with these tools. GenAI has begun to play a role in educational settings, where it accelerates project execution~\cite{zheng_chatgpt_2023, amoozadeh_trust_2024}. Instructors who favor integration wish to use AI for personalized student help, reduce their own time spent on repetitive tasks, and focus more on code critique and open-ended design assignments~\cite{lau_ban_2023}. For example, GenAI has proven helpful in programming courses, where it helps students to overcome learning barriers by providing real-time feedback and assisting in coding tasks~\cite{daun_how_2023}. \citet{kuramitsu_kogi_2023} study indicates that GenAI can be used collaboratively, suggesting and helping students solve coding issues by themselves. Rather than accepting students to use these tools unsupervised, software engineering curricula need to adapt and prepare students for the new reality of work practice.
Similarly, researchers started looking into how to integrate GenAI into Design Sprint activities and support human-AI collaborative technology design~\cite{grangeHumanGenAIValueLoop2024}, and how to facilitate collaborative GenAI image prompting~\cite{hanWhenTeamsEmbrace2024}. Specifically for generative image models, researchers are now building tools and interfaces to help designers generate images with GenAI more efficiently for HCI technology design purposes. For example, tools and interfaces using GenAI to build digital moodboards~\cite{pengDesignPromptUsingMultimodal2024}, storyboards~\cite{liangStoryDiffusionHowSupport2024}, 3D Designs~\cite{liu3DALLEIntegratingTextImage2023} and tools simplifying the prompting experience~\cite{wanGANCollageGANDrivenDigital2023}.

\subsection{Unexplored Challenges: GenAI's Role in HCI Education}

Despite GenAI's growing presence in technology design, there is limited empirical evidence on how it is currently being used by students training to become HCI practitioners. Notable exceptions are a position paper by \citet{maceli_incorporating_2024} exploring students' use of image generation tools for HCI assignments, and research by \citet{kharrufa_potential_2024} alongside a workshop paper by \citeanon{sandhaus_student_2024} both investigating students' use of GenAI in HCI coursework. Their work point to an ethically multi-faceted challenge of integrating GenAI in HCI classes; both from the diversity of use cases it can be applied to, as well as seemingly contradictory uses of GenAI. I.e., \citeauthor{kharrufa_potential_2024} finds students believing GenAI to be both mitigating and re-enforcing the design of biased user personas, as well as being non-creative but using it for creative tasks. While these studies have examined the role of AI in accelerating design practices and offering new opportunities for creativity, they show how little we as HCI educators know about how 1) these tools can be used for HCI tasks, 2) how students use them, and 3) how we should teach students not to use them.

In response to \citet{li_user_2024}, this study aims to unpack the impact of GenAI on novice UX Designers practice. We conclude that the influence of GenAI on this field necessitates prompt adaptation. This paper unravels how students in an interactive device design course can use it to learn less, or more, as well as to design better as well as worse. Through our mixed methods study design we highlight the students' perspective and discuss recommendations for enhancing curricula to prepare future designers for AI-assisted creative work.

\section{Method}
Building upon prior research that highlights the potential advantages of GenAI in HCI design, and concerns to de-skill novice HCI practitioners, we investigate the use of GenAI by students in a course taught by the authors after instruction. In this section, we detail the structure of the HCI course, the original policy implemented regarding GenAI usage, the participant demographics, and the data analysis. In an effort towards transparency~\cite{Wacharamanotham2020-rz}, we make supplementary materials—including the course syllabus, interview protocols, interview transcripts, student-GenAI chat logs, the coding schema for interviews, and students final project designs—accessible upon publication in an Open Science Foundation Repository~\footnote{\label{osf_repo}\url{https://osf.io/wt742/?view_only=8320394c18d04f818c8be45ff57b2636}}. This research was approved as exempt by the Institutional Review Board (IRB) at \textanon{Cornell}{Anonymous} University under \textanon{IRB0148132}{XXXXXXXXX} number, and followed the principles of ethical research conduct outlined by the review board guidelines for human participant research.

\subsection{HCI Class on Design of Interactive Devices} 
\begin{figure}
    \centering
    \includegraphics[width=1\linewidth]{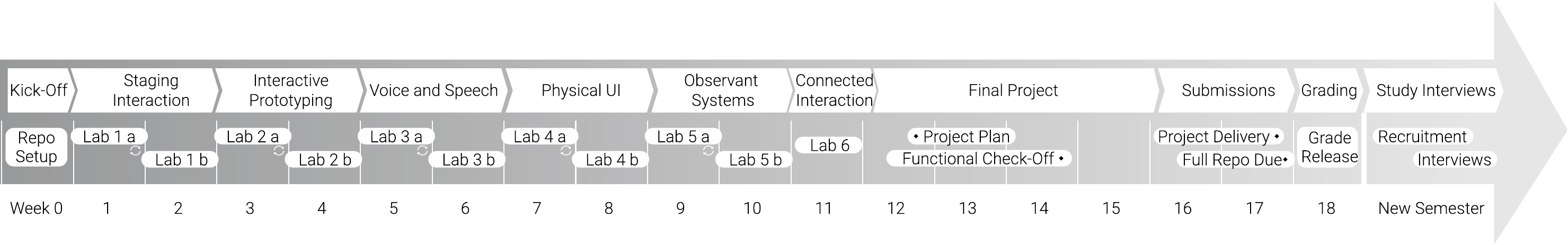}
    \vspace{-20pt}
    \caption{Course Delivery Timeline}
    \label{fig:timeline}    
    \Description{This figure presents a timeline of course delivery for a design course, represented as an arrow pointing to the right. It shows various stages of the course, including Kick-Off and Repo Setup in week 0, followed by Staging Interaction (Lab 1a and Lab 1b in weeks 1 and 2), Interactive Prototyping (Lab 2a and Lab 2b in weeks 3 and 4), Voice and Speech (Lab 3a and Lab 3b in weeks 5 and 6), Physical UI (Lab 4a and Lab 4b in weeks 7 and 8), Observant Systems (Lab 5a and Lab 5b in weeks 9 and 10), Connected Interaction (Lab 6 in week 11), Final Project (Project Plan and Functional Check-Off in weeks 12-14), Submissions (Project Delivery and Full Repo Due in weeks 15-16), Grading (Grade Release in week 17), and finally Study Interviews in the new semester.}
\end{figure}

The \emph{Interactive Device Design} class was a master-level, project-based course focusing on prototyping and interaction design. Forty-eight students iterated through the entire process of designing functional prototypes in two-week sprints, culminating in a final project (see~\autoref{fig:timeline}). The course emphasized hands-on experiential learning, where students were expected to apply user-centered design (UCD) methodologies in a prototyping driven product development process~\cite{soderback2020ux}. The primary learning outcome was to familiarize students with design methods through repeated practice, enabling them to apply these techniques confidently and internalize the principles of interaction design.While the specific implementations may vary, similar courses are offered at most Western universities that have an HCI department.

Additionally, but unplanned, this approach provided students with numerous repeated opportunities to explore various GenAI co-design tactics. 
The syllabus, as well as an overview of lab assignments, are available in an Open Science Foundation repository after publication\footref{osf_repo}.

At the time, the university had not established university-wide policies regarding the use of GenAI in academic settings. As these technologies were only becoming widely available, there was minimal guidance on integrating them into classes and no guidance on using them within the context of user-centered design. Consequently, all student engagements with this evolving technology were exploratory in nature.

\subsection{Class Policy} 
In the \emph{Interactive Device Design} class, the use of GenAI was only permitted with appropriate acknowledgment and advance notice, as detailed by the Class Policy on Academic Integrity (Box ~\ref{box:policy}). Students were not motivated to use GenAI. If they did, the approach was to treat it like any other online reference to avoid plagiarism.

\begin{tcolorbox}[float*=t, width=\textwidth, colframe=gray, colback=gray!10, title=Class Policy on GenAI Usage, sharp corners=south] \footnotesize
\textbf{Academic Integrity}

Each student in this course is expected to abide by the \textanon{Cornell}{Anonymous} University's Code of Academic Integrity. Any work submitted by a student in this course for academic credit must be the student's own work. The policy can be found on the University’s website here: \linkanon{https://gradschool.cornell.edu/policies/academic-integrity/}.

In this class, substantial use of online materials and open-source software is encouraged. You are welcome to utilize found code and online examples, and collaborate as part of an open-source community. However, proper attribution of all work, assistance, and collaboration is essential. 

You must document and notify the teaching team in advance if you plan to use:
\begin{itemize}
    \item Work from other concurrent/prior courses,
    \item Assistance from outside students or people,
    \item Assistance from ChatGPT or other online services.
\end{itemize}

This is permitted, but the net amount of work in this and other courses should be equivalent to pursuing different projects independently. The faculty may impose grade penalties for unattributed copying. Penalties for willful and egregious violations can include course failure and University disciplinary action.
\label{box:policy}
\end{tcolorbox}

\subsection{Participants} 

To investigate students' use of GenAI in HCI education, we interviewed 17 students from the \emph{Interactive Device Design} class, conducting a total of 12 group interviews. The participants comprised a diverse cohort of master's students from various disciplines, including Computer Science, Social Sciences, Design, as well as more distant fields such as Urbanism, Law, and Medicine, exhibiting a range of experience levels in both design and coding. In the recruitements, we aimed to include both students who had used GenAI in their projects and those who had not. With nine of the participants male, we had a roughly equal gender balance. The class had a total of 13 groups, of which 10 groups participated in the interviews. The interview groups background information are detailed in \autoref{tab:participants}. 
\begin{table}[pt]\label{tab:participants}
  \resizebox{\textwidth}{!}{%
  \begin{tabular}{@{}c|ccc|cccccc@{}}
    \toprule
  \begin{tabular}[c]{@{}c@{}}Interview \\ Nr.\end{tabular} &
    \begin{tabular}[c]{@{}c@{}}Group \\ Size\end{tabular} &
    \begin{tabular}[c]{@{}c@{}} Experienced \\ Coder\end{tabular} &
    \begin{tabular}[c]{@{}c@{}} Experienced \\ Designer\end{tabular} &
    \begin{tabular}[c]{@{}c@{}}Participants \\ Present\end{tabular} &
    \begin{tabular}[c]{@{}c@{}}Previous \\ GenAI \\ Experience\end{tabular} &
    \begin{tabular}[c]{@{}c@{}}GenAI \\ Use\end{tabular} &
    \begin{tabular}[c]{@{}c@{}}GenAI \\ Attribution \vspace{-3pt}\\ \tiny (Out of \$1000)\end{tabular} \\ \midrule
  \textbf{1}  & 5 & Yes & Yes  & 3 & High      & High & \$166  \\
  \textbf{2}  & 6 & Yes & Yes  & 1 & Low      & Low & \$10  \\
  \textbf{3}  & 4 & Yes & No   & 2 & High      & Low & \$200 \\
  \textbf{4}  & 6 & No  & Yes  & 1 & Low      & High & \$20  \\
  \textbf{5}  & 2 & No  & Yes  & 2 & High      & High & \$75  \\
  \textbf{6\textsuperscript{*}}  & 6 & Yes & Yes  & 1 & Low      & High & \$200 \\
  \textbf{7\textsuperscript{*}}  & 6 & Yes & Yes  & 1 & Low      & Low & \$1   \\
  \textbf{8}  & 3 & Yes & Yes  & 1 & Low      & Low & \$40  \\
  \textbf{9\textsuperscript{†}}  & 2 & Yes & No   & 1 & None & Low & \$350 \\
  \textbf{10} & 2 & No  & Yes  & 2 & High      & High & \$200 \\
  \textbf{11} & 1 & Yes & No   & 1 & Low      & Low & \$200 \\
  \textbf{12\textsuperscript{†}} & 2 & Yes & No   & 1 & Low      & Low & \$200 \\ \bottomrule
  \end{tabular}%
  }
  \caption{Participant information on group composition with prior experience (left), and interview insights including, GenAI use, and attribution of a hypothetical group award split among the project group (right).  Experienced coder and designer refers to the presence or absence of students in the group with prior experience in these practices. Asterix\textsuperscript{*}, and dagger\textsuperscript{†} indicate interviews with students from the same final groups.}
  \Description{This table provides information on participant group composition, previous experiences with GenAI, and how much hypothetical attribution was given to GenAI out of a total of \$1000. The table includes columns for interview number, final project group size, whether the group had an experienced coder or designer, the number of participants present during the interview, previous GenAI experience, GenAI use, and the attributed value to GenAI in dollars. Group sizes vary from 1 to 6, and GenAI attribution ranges from \$1 to \$350. Some interviews are from the same final groups, indicated with an asterisk or dagger symbol.}
  \end{table}

\subsection{Interview Process}
Two researchers conducted the semi-structured interviews. The interviews were conducted post-course after the students had received their final grades in the following semester. All interview participants were informed that their responses would remain anonymous and that disclosing the use of GenAI would not affect their academic standing, even in cases of cheating and academic dishonesty. 
These interviews explored what students used GenAI for (\autoref{ssec:what}), how students used GenAI~(\autoref{sssec:usage-taxonomy}), and students' reflections on GenAI's implications for user-centered design, prototyping, and education~(\autoref{ssec:why}). The interview script is available in the Open Science Foundation\footref{osf_repo}. 

Participants were asked to share their screen in the online interviews and have their class repository and GenAI tools, including their Chat History, open. The interviewer reviewed all class submissions with the interviewees to prompt recollections of their GenAI usage. By inquiring about specific characteristics indicative of ChatGPT-generated content—such as verbose language, excessive reliance on itemized lists, and bolded introductory phrases (which were distinctive to ChatGPT creations at that time~\cite{krissydavisHowTellIf2023}) —, the interviewer pushed participants to openly discuss their experiences with GenAI. Interviews were video recorded, transcribed afterward using a local transcription service (OpenAI Whisper~\cite{OpenaiWhisper2022}), and manually anonymized before analysis.

\subsection{Interview Analysis}
We used a multi-stage coding process to analyze the interviews. This process involves three rounds: initial inductive coding in two rounds and a final round of deductive coding, including manual sentiment analysis . We based our approach on a multi-cycle coding process outlined by \citet{Vanover2021-xm}).
\begin{enumerate}
    \item \textbf{First Round (Inductive Coding by Interviewers):} During the interviews, one researcher took notes while the other conducted the interview. Both researchers attended all interviews, with one rewatching the sessions they missed. An inductive round of rough coding was performed, focusing on identifying surprising uses of AI and capturing students' reflections on what went well and what did not.
   \item \textbf{Second Round (Independent Inductive Coding):}
   A third researcher, who did not conduct the interviews, performed another round of inductive coding. They began by reading the transcripts and identifying broader themes while having access to the initial round of codes. They performed thematic analysis by delving deeper to find smaller codes and grouping them into themes.
   \item \textbf{Third Round (Codebook Application):} In the last coding round, the established codebook was used deductively on all interview transcripts, and coding disagreements were resolved. The full list of high level themes were: AI use purposes and scenarios, Advantages and positive impacts of AI, Limitations and challenges of AI, Comparison of AI with traditional methods, and The implications and ethical considerations of AI in education.\end{enumerate}

After coding the interviews, we applied journey mapping~\cite{endmannUserJourneyMapping2016} using the established codebook theme 'AI use purposes and scenarios' to match students' GenAI use to the UX process, and affinity diagramming~\cite{rachelkrauseAffinityDiagrammingCollaboratively2024} using the other themes to analyze them for findings. We used a mixed methods approach to quantify sentiment categories and provide rich qualitative descriptions.

The transcripts, finalized codebook, interview excerpts linking GenAI use cases to the UX design cycle, and an affinity diagram are publicly available in an Open Science Repository~\footref{osf_repo}. Interviewers captured screenshots of student interactions with GenAI, and many students provided their own interactions through PDFs, screenshots, and chat links. This supplementary data was analyzed for insights into student engagement with GenAI. The GenAI Interaction Samples will also be archived in the Open Science Repository~\footref{osf_repo}.

\paragraph{Positionality and Authorship} %
The authors are HCI researchers and educators with expertise in interaction design, qualitative research, and prototyping. The first author is a Ph.D. student with a UX background, the second is an international undergraduate researcher, the third is a Ph.D. student skilled in HCI design and machine learning, and the last is a faculty member experienced in HCI and design.
While some authors taught the \emph{Interactive Device Design} class, the second author provided an external perspective. The authors acknowledge that their roles may have influenced the study and aim for transparency through detailed research methods and open data sharing.

Authors used GenAI to correct grammar, proofread, and assist with LaTeX formatting but claim full authorship~\cite{ACM-Publications-Board2023-wp}. GenAI was not used for data analysis, findings review, or drafting.  %

To conclude the method section, we have detailed the structure of the HCI course, the original policy implemented regarding GenAI usage, the groups compositions, and the methodologies employed for data analysis. In the next section, we present the findings from the interviews, detailing the students' experiences with GenAI in HCI education focusing on the questions of what they used GenAI for how, they made GenAI useful for their HCI tasks, and why they think GenAI is appropriate or not appropriate to use in HCI practice and education.

\section{Results}
We find that students attempt to use GenAI for most interaction design tasks assigned to them. We begin by briefly describing these design and prototyping tasks, highlighting how students applied GenAI within the interactive device design education class~(\autoref{ssec:what}). This includes examining the specific ways students interact with GenAI, the patterns of use that emerged~(\autoref{sssec:usage-examples}), as well as a taxonomy of usage patterns~(\autoref{sssec:usage-taxonomy}).

Next, we discuss students' sentiments and feelings about GenAI~(\autoref{ssec:why}), including instances where GenAI was either successful or fell short in aiding their work. This section also explores their emotional reactions to using the technology and its perceived effectiveness in different contexts.

Finally, we present students' reflections on the broader implications of GenAI in the HCI design process~(\autoref{sec:impact}), focusing on their views of GenAI’s potential role in shaping future HCI design practices.

\subsection{What Students' Use GenAI for in Interactive System Design}
\label{ssec:what}

We found that all participants had used GenAI in their projects, with varying degrees of engagement, but more extensively than disclosed through class submissions.
Students reported experimenting with various GenAI tools for text generation, such as Grammarly AI\footnote{\href{https://www.grammarly.com/ai}{grammarly.com/ai}}, Google Bard\footnote{\href{https://bard.google.com/}{bard.google.com}}, Bing Chat\footnote{\href{https://chat.bing.com/}{chat.bing.com}}, and Copilot\footnote{\href{https://github.com/features/copilot}{github.com/features/copilot}} for coding support, as well as for image generation, including Midjourney\footnote{\href{https://www.midjourney.com/}{midjourney.com}}, Adobe Firefly\footnote{\href{https://www.adobe.com/products/firefly.html}{adobe.com/products/firefly}} Google Labs AutoDraw\footnote{\href{https://www.autodraw.com/}{autodraw.com}}, and DALL-E\footnote{\href{https://openai.com/dall-e-2}{openai.com/dall-e-2}}. However, ChatGPT versions 3.5 and 4.0\footnote{\mbox{\href{https://openai.com/research/gpt-4}{openai.com/research/gpt-4}}} were by far the most frequently used tools, with few competitors.

We identified 9 main use cases throughout the HCI prototype design process for which students used GenAI tools:

\begin{itemize}[leftmargin=*,itemsep=0pt,parsep=1pt, label={$\diamond$}]   
    \item \textbf{Brainstorming:} GenAI was leveraged for generating ideas for technology applications, either as a starting point for prototype planning or as additional input.
    
    \item \textbf{Use Case Development:} GenAI assisted in developing user personas and synthesizing user characteristics, needs, and goals.

    \item \textbf{Storyboard generation:} Students used GenAI to create storyboards, visualizing user scenarios, journeys, or interactions, aiding in the definition and planning of the UX/UI design process.

    \item \textbf{Asset generation:} Tools like DALL-E were utilized for creating visual assets or mockups, supporting the design phase by generating images or icons. Additionally, students produced images for interactive prototypes, including background visuals and status indicators.  Students also used GenAI for copywriting of textual content, such as outputs for speech prototypes and user feedback messages for the interactive prototypes.

    \item \textbf{Prototyping Coding Assistance:} Tools like ChatGPT were used for coding prototypes, from generating initial code snippets to providing debugging solutions and refining prototypes.
    
   \item \textbf{Hardware Integration:} Students used GenAI to solve hardware problems, including sensor detection and compatibility and cabling.
   
   \item \textbf{Documentation and Writing Assistance:} GenAI was used for automating the creation of project documentation by cleaning up text, expanding notes, and adding code comments.

\item \textbf{Design Reflection:} GenAI was applied to synthesize and analyze user feedback or test results. Students also used it to reflect on their own work process, as well as other students' work.

\item \textbf{Learning Support:} Students reported using GenAI for explaining concepts, deriving instructions, and learning about topics generally, not limited to the context of prototype development.

\end{itemize}
\begin{figure}[t]
    \centering
    \subcaptionbox{Generating text for an overall storyboard narrative (I9, left) and descriptions for individual frames (I5, right).}{%
    \includegraphics [height=0.250\textwidth]{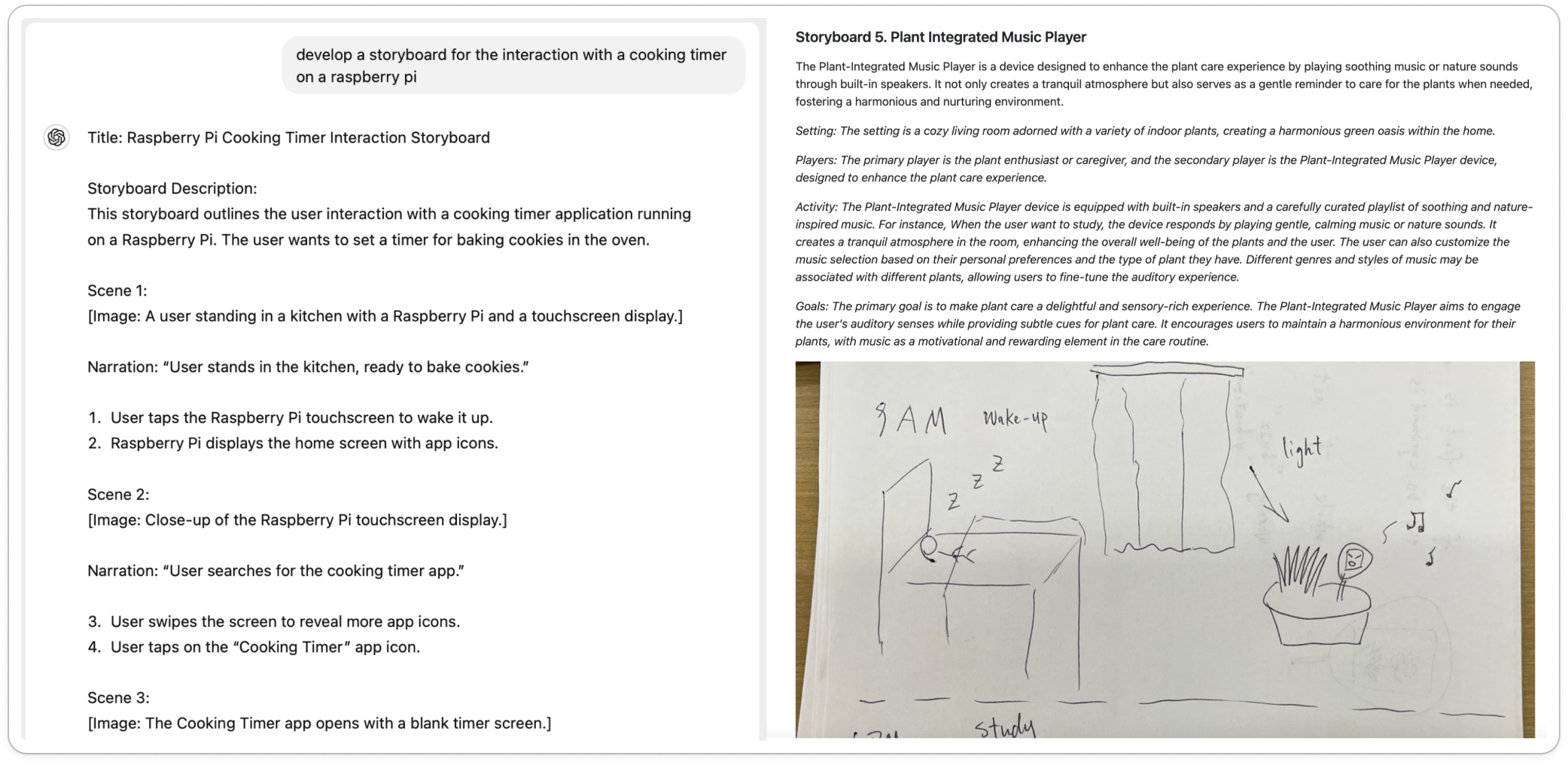}} \hspace{10pt} \vspace{10pt}
    \subcaptionbox{Generating individual assets to compose into storyboards: icons (I9, left) and styled image assets (I10, right).}{%
    \includegraphics[height=0.250\textwidth]{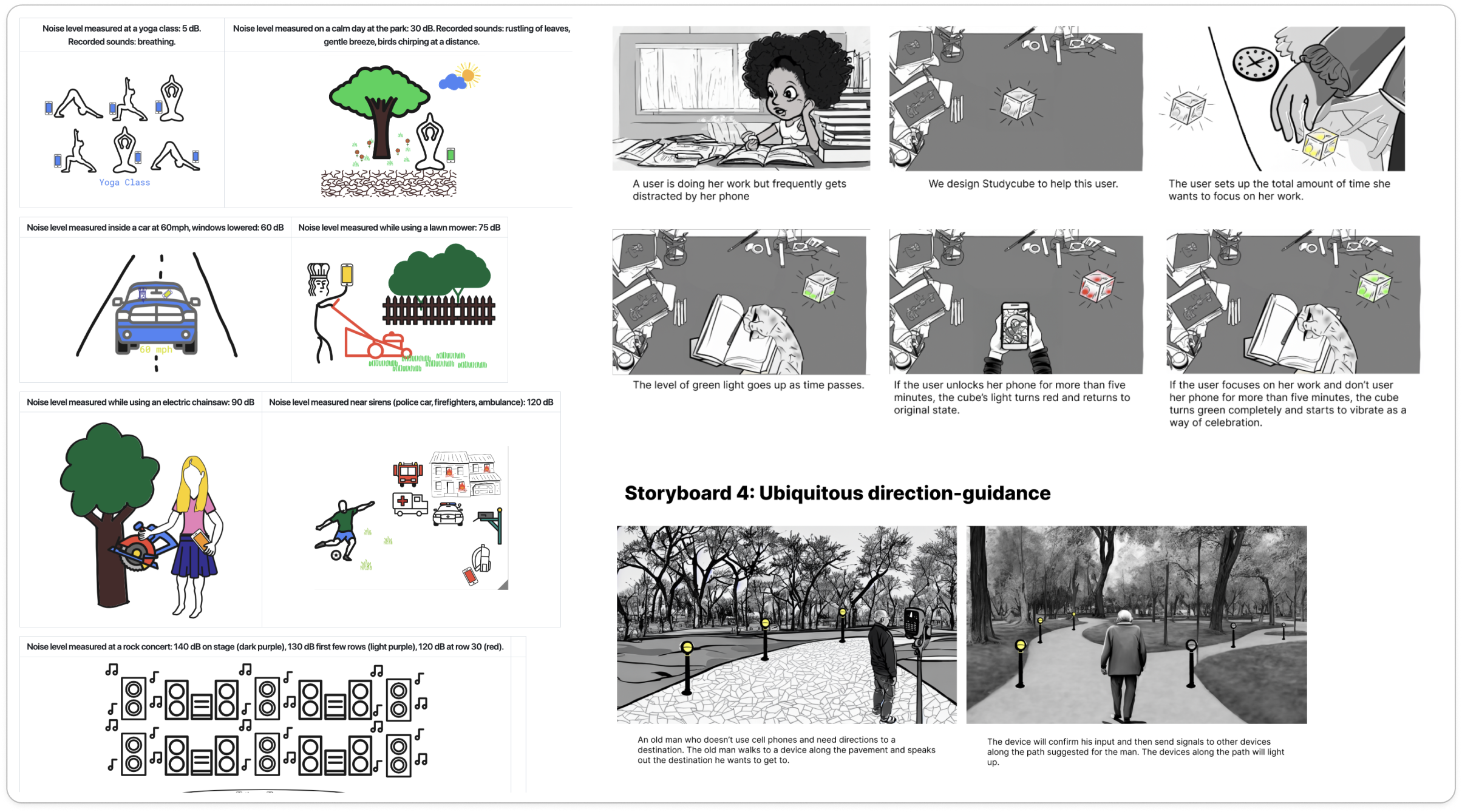}}     
    \subcaptionbox{Generating individual frames one by one (I1).}{%
    \includegraphics[height=0.182\textwidth]{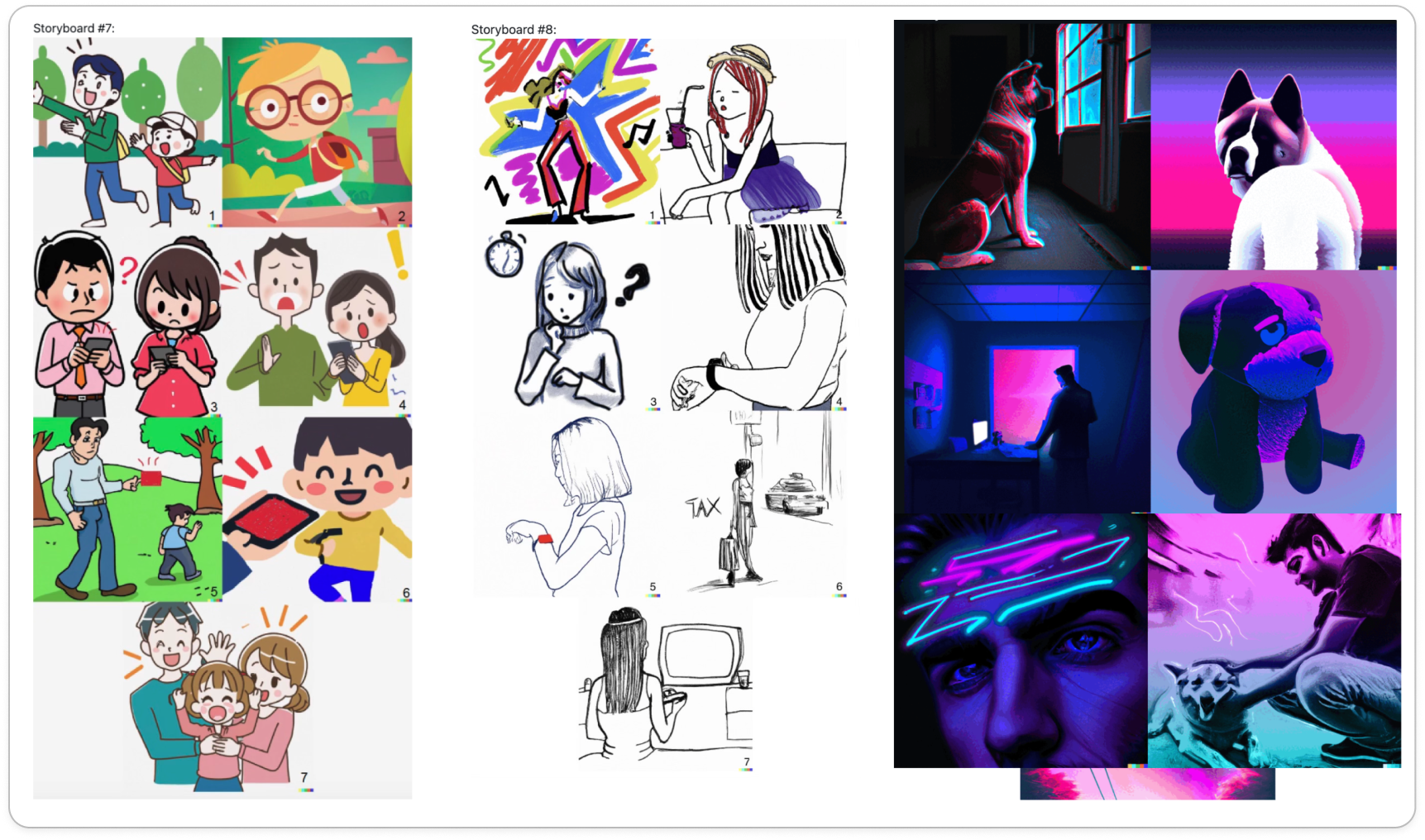}} \hspace{10pt}
    \subcaptionbox{Generating background frames with device and user interactions drawn on top manually (I1, I3).}{%
    \includegraphics[height=0.182\textwidth]{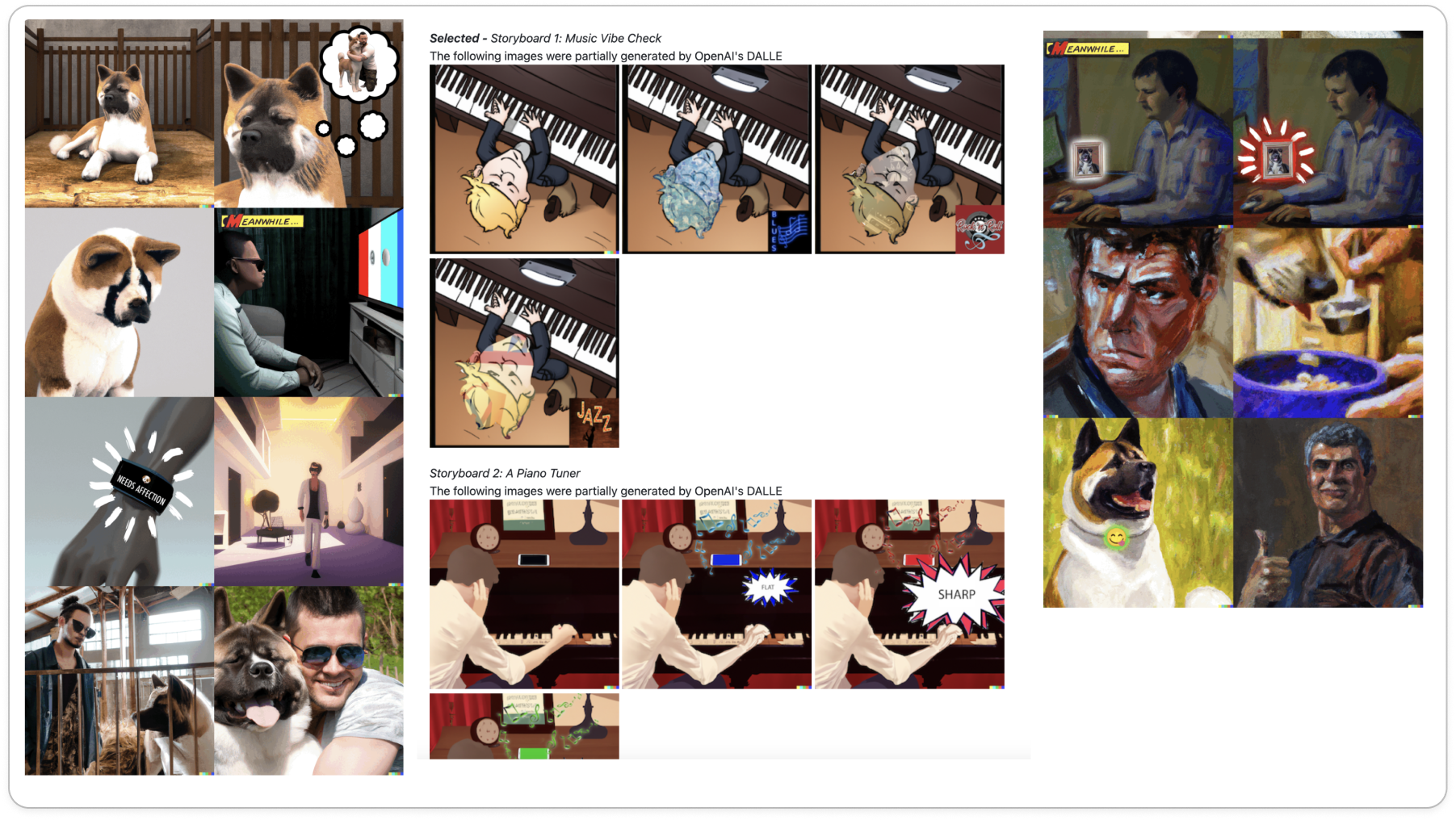}} \hspace{10pt}
    \subcaptionbox{Generating the entire storyboard at once (I5).}{%
    \includegraphics[height=0.182\textwidth]{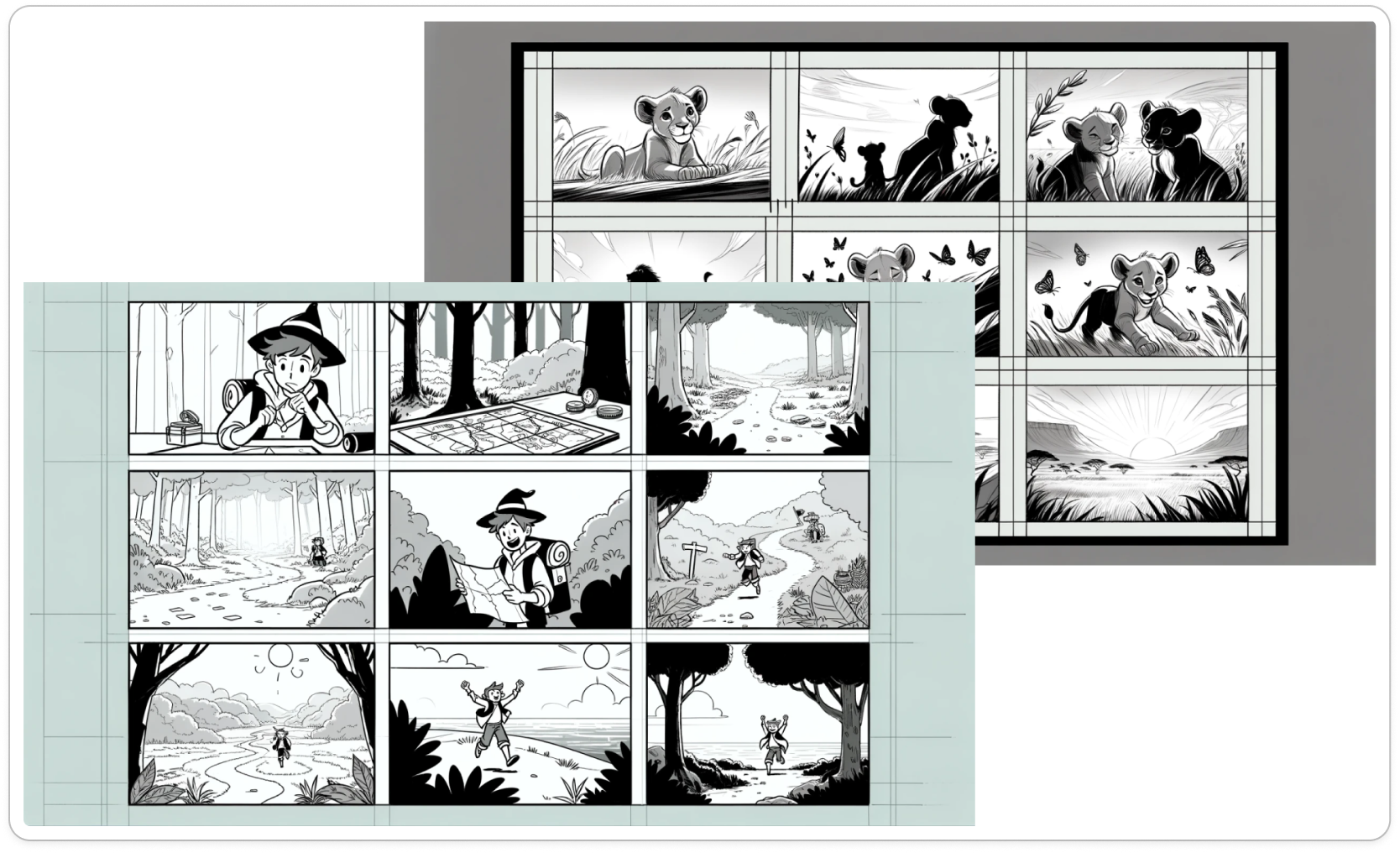}}

    \caption{Students take diverse strategies when interacting with GenAI for storyboarding.}
     {\footnotesize Image credit: \href{https://github.com/jamiewang76/Interactive-Lab-Hub/tree/Fall2023/Lab\%201}{Jiao and Wang}, \href{https://github.com/annetta-zheng/Interactive-Lab-Hub/tree/Fall2023/Lab\%201\#lab1-a}{Zheng}, \href{https://github.com/wjr83/Interactive-Lab-Hub/tree/Fall2023/Lab\%201}{Bhansali and Reid}, \href{https://github.com/nojcc/Interactive-Lab-Hub/tree/Fall2023/Lab\%201}{Aarons, Bhigroog, Caceres, Minkowitz and Xu}, \href{https://github.com/zacharypakin/Interactive-Lab-Hub/tree/Fall2023/Lab\%201}{Suberviola, Nikitovic and Patil}, \href{https://github.com/sophiewu7/Interactive-Lab-Hub/tree/Fall2023/Lab\%201}{Wu and Sun}}\label{fig:genai_storyboard_behaviors}
    \Description{
        The figure showcases various strategies students employed to use GenAI in storyboarding, with five distinct image groupings. 
        (a) Shows a text-based storyboard narrative on the left and descriptions for individual storyboard frames on the right.
        (b) Displays individual assets generated by GenAI, with icons on the left and styled image assets on the right.
        (c) Depicts individual storyboard frames generated one by one, showing a sequence of images in different styles.
        (d) Presents a background frame generated by GenAI, with user interactions and technology design details manually added.
        (e) Shows a full storyboard generated by GenAI, with all frames produced at once, arranged in a consistent style and simple storyline.
        }
    \end{figure}

The majority of groups used GenAI for prototyping  (9/10) and brainstorming (7/10). Around half the groups used it for asset generation (6/10), hardware-related problems (6/10), documentation (5/10), design reflection (5/10), and learning (6/10). The least used tasks were use case development (4/10) and storyboard generation (4/10). \autoref{fig:use-case-examples} provides illustrative examples of these applications, which can be mapped onto the Double Diamond Design Process (appendix \autoref{fig:double-diamond}).
\subsubsection{\textbf{Usage Patters across Use Cases}}

As students ran through the design thinking and prototyping assignments multiple times, they had multiple chances to try GenAI, and over time, usage patterns changed. 
From the groups that used GenAI for brainstorming a majority of participants (4/7) reported to have stopped the use for it. Additionally, half (2/4) of the groups either reduced or ceased their usage for storyboard generation, while a minority (2/6) stopped using it for asset generation. Notably, only one group (1/9) discontinued its use for prototyping, whereas two groups increased their utilization of GenAI for this purpose. No significant changes in usage were observed for use case development, hardware assistance (though 1 group reduced their usage), as well as for learning support, documentation assistance, and design reflection. Detailed information on the use case changes by the groups can be found in the OSF repositories HCI Tasks Use Case table~\cite{Sandhaus_Gu_Parreira_Ju_2024}.

In analyzing these usage patterns, we consider not just \textit{what} students used GenAI for but also \textit{how} they approached these tasks. Below, we leave examples of the range of co-design strategies that students employed to achieve specific goals within prototyping design.

\paragraph{\textbf{Storyboard Generation:}}
Some groups used ChatGPT to generate text narratives for the overall storyboard, whereas others created textual descriptions for individual frames~(\autoref{fig:genai_storyboard_behaviors} a). In terms of visuals, some students generated individual visual assets such as icons or styled images that they arranged then manually ~(\autoref{fig:genai_storyboard_behaviors} b), or full storyboard frames one by one~(\autoref{fig:genai_storyboard_behaviors} c). Certain teams combined GenAI-generated background frames with manually drawn details~(\autoref{fig:genai_storyboard_behaviors} d), and some generated the entire storyboard at once~(\autoref{fig:genai_storyboard_behaviors} e).

\paragraph{\textbf{Prototyping Coding Assistance:}} 
GenAI was very commonly used for debugging and reorganizing code \inlinequote{GPT is just a better version for debugging, a better tool.}{10}, replacing online coding support and library documentation \inlinequote{[When] I couldn't find like a proper documentation quickly, I would just ask the questions to check [GPT] on how to call this rather than start to Google and then try and find Stack Overflow or community answer questions.}{8}, as well as adapting examples from the web \inlinequote{When the provided commands from tutorials or Stack Overflow are overly complex or not suited to my system's specifics, I use ChatGPT to simplify them.}{5}. 

But also included in many part of technology prototypes to simulate, final functionality such as Application Programming Interface (API)  \inlinequote{We used the ChatGPT AI API playground to make the speech input more polite [and] ensuring our device communicated in a friendly manner.}{4}, and setting up initial code structure to prototype from \inlinequote{Asking Bard, hey, can you give me like a basic structure for a mastermind game. [...] I was able to get a working solution for it.}{3} or prototype without coding at all \inlinequote{For some teammates, they prefer to do prompt engineering, so they will just, you know, [when assigned a task] they just reiterate the task into prompts and then ask the chatbot to write the code for them.}{6}.

\paragraph{\textbf{Hardware Integration Assistance:}} 
\label{sssec:usage-examples}
\begin{wrapfigure}{R}{0.35\textwidth}
    \centering
\includegraphics[width=0.34\textwidth]{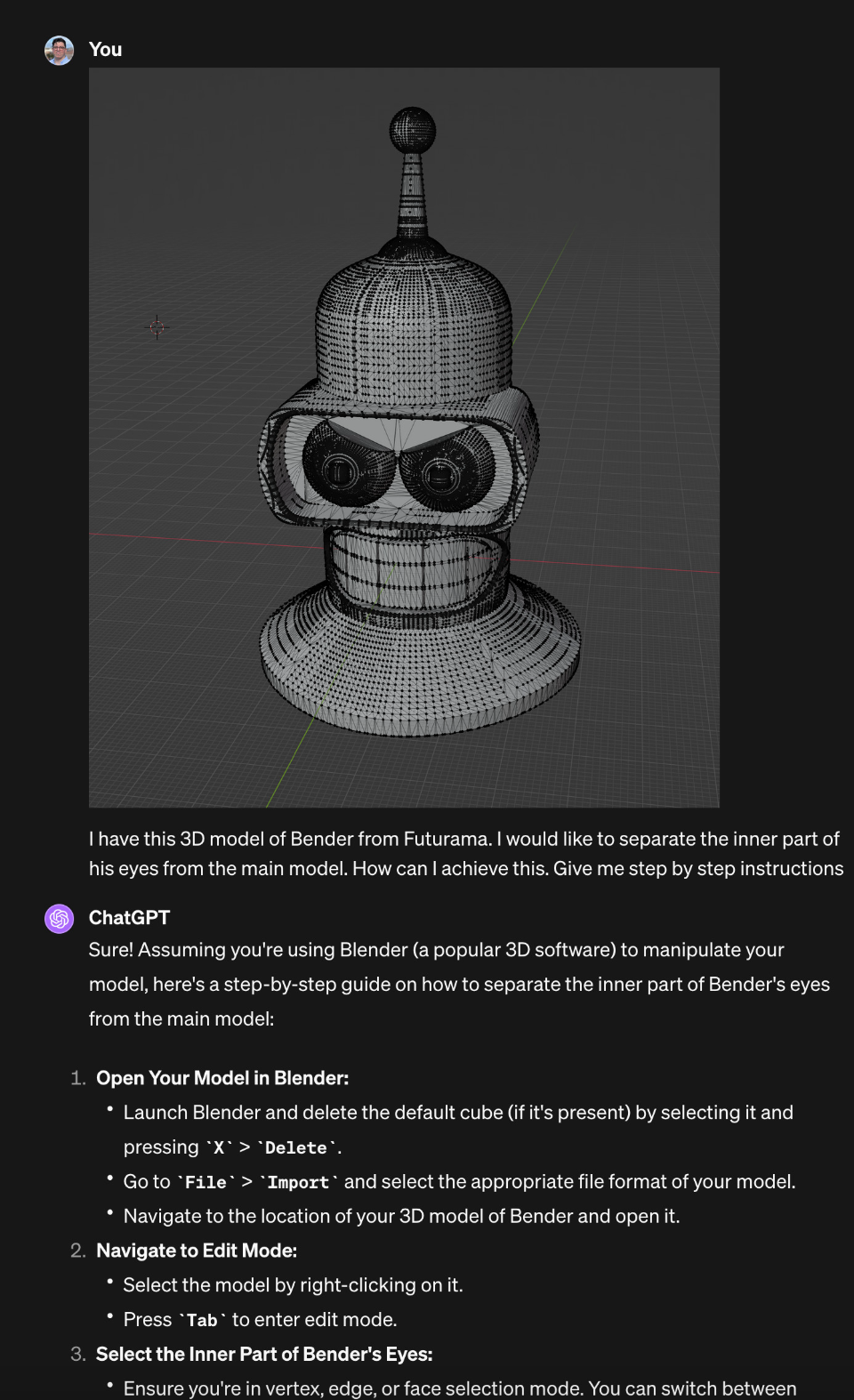}
    \caption{3D model of Bender from Futurama, with hardware integration instructions generated by ChatGPT (I1).}
    \label{fig:blender}
    \Description{A screenshot from a 3D model of Bender from Futurama, with hardware integration instructions generated by ChatGPT.}
\end{wrapfigure}
Students used GenAI for hardware prototyping assistance, students use it as an initial step to understand the hardware capabilities. One student noted, \inlinequote{When it comes to new hardware, we always [consult] GPT first. We gave it the exact model we had and then asked how to incorporate that into our Python project on the Raspberry Pi.}{10}. Additionally, they sought guidance on interpreting and integrating various sensors, with one remarking, \inlinequote{[It] helps to explain incompatibilities between certain types of sensors, and with the functionality of the sensor when we didn't know how to use it.}{12}. Furthermore, students received instructions on 3D modeling mechanical components using design and printing software, as illustrated by a student's experience: \inlinequote{It helped me learn how to remove certain assets from a model. [...] We had to print them in separate pieces so we could fit all the different hardware inside of him, [...] a camera inside of his mouth [...] lasers behind his eyes. The model I imported into Blender was one object. I had to use GPT to understand how to separate everything so that I could print all those different pieces separately with a 3D printer.}{1} (see~\autoref{fig:blender}).

\subsubsection{{\textbf{A Taxonomy of Student GenAI Usage Patterns}}}\label{sssec:usage-taxonomy}

In analyzing how students interact with GenAI in their HCI design process, four primary patterns emerged: \textbf{Benchmark}, where GenAI helps students compare and improve their work; \textbf{Booster}, assisting in overcoming challenges; \textbf{Executor}, following instructions to complete tasks; and \textbf{Amplifier}, where AI generates results beyond students' capabilities.

AI usage was more varied in creative and reflective tasks compared to execution-focused ones. In the Discover, Define, and Reflection stages, students applied GenAI in diverse ways to explore, conceptualize, and reflect on their projects. For example, during brainstorming, groups highlighted multiple AI roles: Benchmark, \inlinequote{We used GPT for brainstorming but had many good ideas ourselves,}{1} Booster, \inlinequote{AI guided us through problem spaces and suggested overlooked ideas}{3} and Executor, \inlinequote{We used ChatGPT to generate ideas for Raspberry Pi development.}{9} In contrast, AI usage during the Develop stage was more uniform, with all teams exclusively using it as an Executor (6/6) for documentation assistance, as an example. %

\subsection{Students’ Sentiments About GenAI} \label{ssec:why}

Overall, students' sentiment about GenAI appeared to be mixed. In the Definition phase (Brainstorming, Use Case Development, Storyboard Generation), 50\% of students expressed negative sentiment, while the remaining 50\% expressed positive sentiment regarding GenAI's application for tasks. In contrast, during the Execution phase, 46\% of students had a positive sentiment, 36\% had a negative sentiment, and 18\% were uncertain about GenAI's performance in the use cases matched to this phase. When considering GenAI for Learning Support (as its own phase), students' sentiments were leaning towards the negative: 44\% expressed negative sentiment, 33\% were positive, and 22\% were uncertain about GenAI’s role in supporting their learning outcomes.

\begin{figure}
    \centering
    \includegraphics[width=0.97\linewidth]{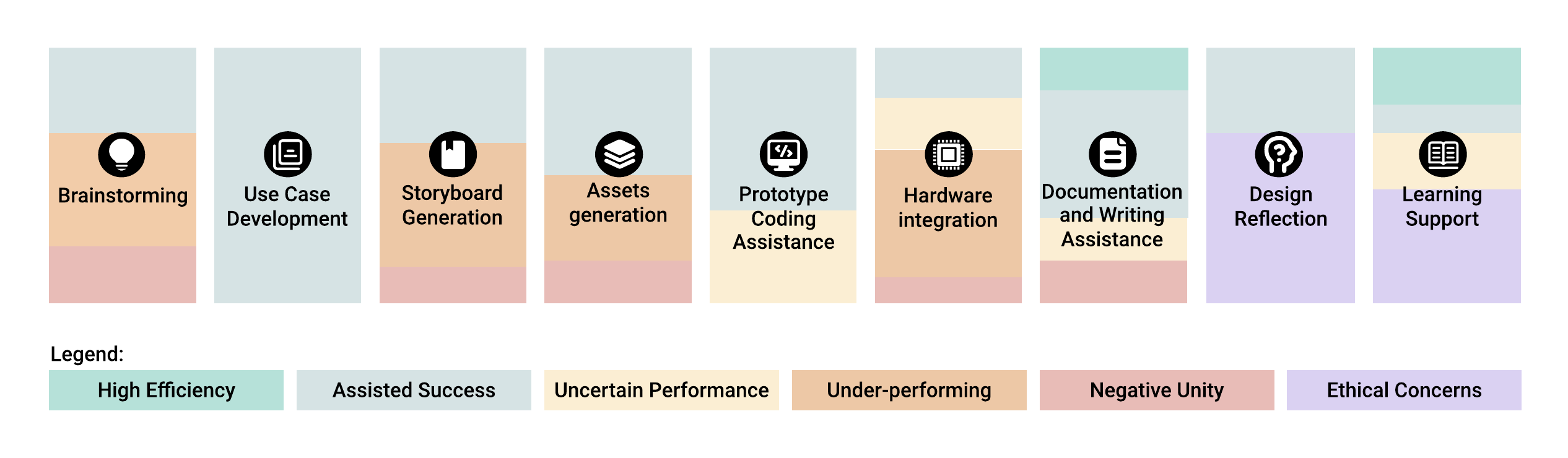}
    \caption{Student's sentiment about each GenAI use case. Data in proportion to reported sentiments.}
    \label{fig:sentiment-overview}
    \Description{This figure presents an overview of student sentiment analysis regarding various use cases of GenAI in an HCI design class. The figure is divided into eight categories: Brainstorming, Use Case Development, Storyboard Generation, Asset Generation, Prototype Coding Assistance, Hardware Integration, Documentation and Writing Assistance, Design Reflection, and Learning Support. Each category is represented by a vertical bar with colors indicating different sentiments, including High Efficiency, Assisted Success, Uncertain Performance, Under-performing, Negative Utility, and Ethical Concerns. A legend is provided at the bottom of the figure to explain the color-coding of each sentiment. Use Case Development Prototype Coding Assistance, appear in more green positive colors, whereas Brainstorming, Storyboard Generation, and Hardware Integration are more red negative colors. Design Reflection and Learning Support are more purple colors showing ethical concerns.}

\end{figure}

From the interview analysis, we code students' sentiments into 6 main categories, both regarding self-reported efficiency of the tools and general ethical sentiment. The sentiment distribution of students for each of the use cases described in \autoref{ssec:what} can be found in \autoref{fig:sentiment-overview}. %
These sentiment groups, emerging from the interview coding, are as follows:
\vspace{5pt}

\noindent\fcolorbox{HighEfficacy}{HighEfficacy}{High Efficacy:} GenAI demonstrates high efficacy in task execution, meeting student expectations with satisfactory results.

Only for Documentation Assistance and Learning Support, students expressed High Efficacy. In these mentions, GenAI meets students expectations with satisfactory results. %
\inlinequote{GPT actually gives you a very detailed explanation on the code that it changed and altered[...].Sometimes after you put in your code, it gives you documentation of the code and that can help give you a clearer picture.}{10}

\noindent\fcolorbox{AssistedSuccess}{AssistedSuccess}{Assisted Success:}  GenAI is beneficial with human intervention, serving as a valuable foundation or supplementary tool.

The most common sentiment across all use cases was "Assisted Success," where students found GenAI to be beneficial with human intervention. %
Assisted success was reported across diverse tasks, such as storyboarding, where students used GenAI-generated text descriptions as a starting point for their own work: \inlinequote{ChatGPT would just give you text descriptions of it, and we would pick the ones that we think are good and then draw it out.}{10}. It was also frequently reported in prototyping, where students acknowledged the tool's limitations but still appreciated its value: \inlinequote{There was a lot of cursing going on, a lot of yelling with GPT, but again, there is nothing else out there like it; it helped us understand the math involved because there was a lot of math when it came to rotating the clock.}{1}. Though refinement was needed, human guidance greatly improved its utility, indicating that the success of GenAI in HCI tasks depends more on how students interact with the tool as detailed in \autoref{sssec:factors}.

\noindent\fcolorbox{UncertainPerformance}{UncertainPerformance}{Uncertain Performance:}  GenAI exhibits mixed performance, providing useful assistance in some areas while falling short in others.

Users experience a combination of satisfaction and frustration, finding the tool helpful for specific tasks but inadequate for others. This sentiment is most prevalent in the prototyping and hardware integration use cases where students often struggle to understand what GenAI could help with, how to prompt it, how to provide it the necessary context to generate useful outputs, or how to interpret it. 
\inlinequote{But a lot of times it's a very good starting point. [...] when you guys gave us new cameras or new sensors to use [...], we didn't know how to start with, what packages to even call that kind of a thing.; let's say the servo motors, you need to set angles, right? [However it] gave us the most trouble with the servo. We could not get that thing to work accurately. GPT kept giving us the wrong instructions on how to fix that. I think I actually took over.}{1}

\noindent\fcolorbox{Underperforming}{Underperforming}{Under-performing:} GenAI exhibits average performance, often failing to meet student expectations.

The tools are deemed by students not to be the optimal choice for task completion. %
 Students often mention that manual work or other traditional tools are faster, more accurate, or more pleasurable. 
\inlinequote{I believe that I prefer Grammarly over ChatGPT. It is not only because Grammarly is faster.}{6}, students do not know how to use GenAI for such a task\inlinequote{It never really occurred to me to use it to draw inspiration from. I'm not sure if that's because I'm not good at prompting.}{8}, or that human collaboration is preferred
\inlinequote{It is just more fun to think about [storyboarding] with the partner.}{12}.

\noindent\fcolorbox{NegativeUtility}{NegativeUtility}{Negative Utility:} GenAI introduces additional complexity and issues during use, with students expressing dissatisfaction with its outputs.

The tool complicates task execution, leading to increased frustration and inconvenience. In these cases, students feel as if they have wasted significant time and effort on GenAI, with little to no benefit.
\inlinequote{I did this I tried using first Adobe, which has a generative AI tool to basically you give it text and it will paint something for you. I [stopped using it] because it [didn't] work well. Later I found a simpler [tool]. Basically, that's the one I used to create these storyboards, but it's still [true that] it would have been so much faster if I would have just drawn it by hand honestly.}{9}

\noindent\fcolorbox{QuestionableIntegrity}{QuestionableIntegrity}{Ethical Concerns:} Students expressed this sentiment primarily in relation to \emph{Learning} throughout the course and \emph{Design Reflection}. They speculated that relying on GenAI could undermine understanding of the design process and the principles of human-centered design. While GenAI can produce work of seemingly high quality, students voiced concerns that such work could be misleading, particularly to graders. Although some positive sentiments emerged in these categories, students generally perceived GenAI as too prone to misuse in learning contexts and in reflecting on design outcomes. This sentiment contrasts with concerns about GenAI’s collaboration efficiency, as ethical issues were noted in both high- and low-efficiency applications. These concerns are explored further with specific examples in \autoref{ssec:Ethical Concerns}.

\vspace{5pt}

The corresponding sentiment quotes for each use case per interview can be found in Appendix \autoref{tab:sentiment-use-cases}.

\subsubsection{\textbf{Factors Influencing GenAI Sentiment and Usage}
\label{sssec:factors}}

Overall, the sentiment classification shows how students' usage patterns modulate their experience with GenAI for interactive device design. Students' stance towards GenAI in HCI design education are complex and multifaceted. While the tool is generally viewed positively, its utility varies across different tasks and phases of the design process; students' sentiments are tremendously influenced by the individual expectations and goals, as well as the specific GenAI tool or model at hand, as well as their prior experience with the specific design task. Evaluation of the tool's performance is also often not homogeneous across groups. We conjecture that patterns of use and use change over time correlate with the perceived efficiency of the tool. We identify three factors which impact sentiment and usage patterns:

\paragraph{\textbf{Experience}}

\begin{figure*}[h]
    \centering
    \includegraphics[width=\textwidth]{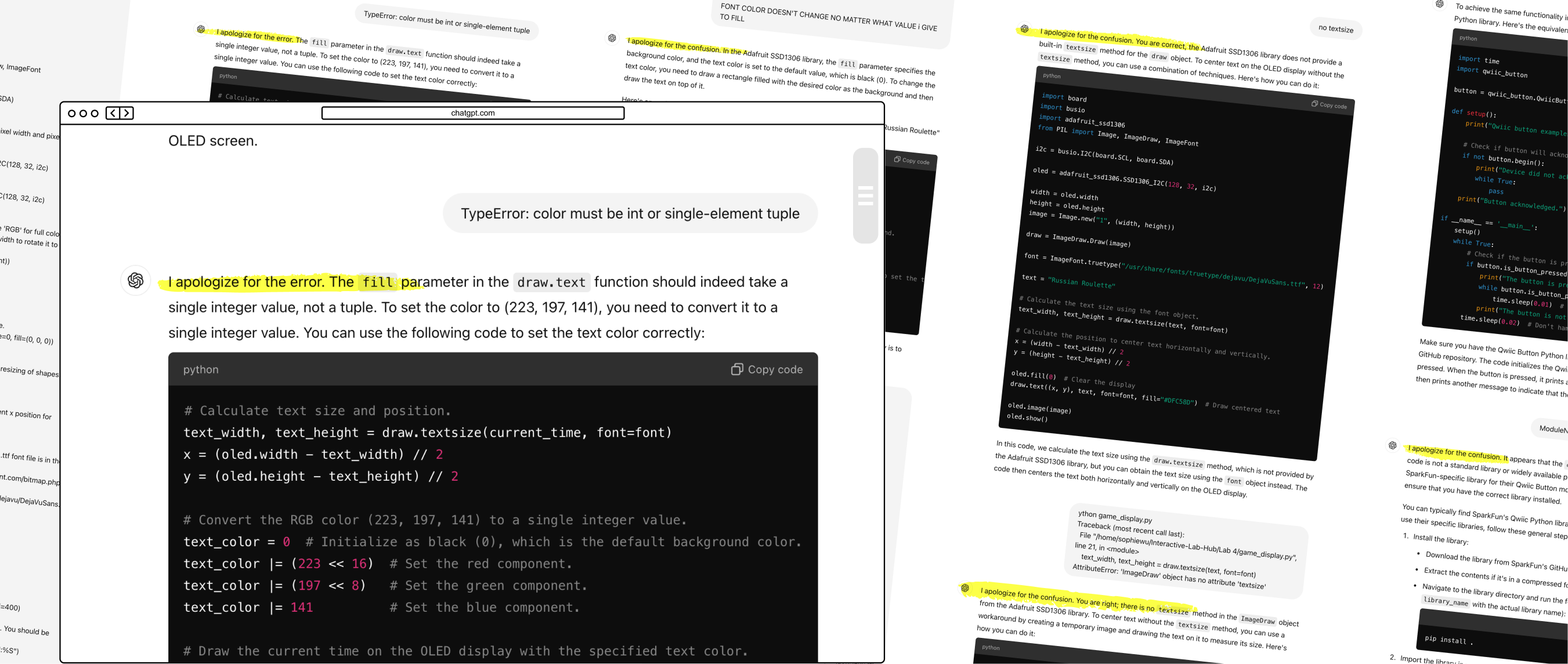}
    \caption{An interaction illustrating a student's struggle with repeated GenAI prompting. The student attempts to resolve a bug by repeatedly submitting the complete code and associated error messages to a GenAI tool, ultimately failing to address what is revealed to be a minor issue (I9). More GenAI interactions archived on OSF~\cite{Sandhaus_Gu_Parreira_Ju_2024}.}
    \label{fig:prompt-failure}
    \Description{The figure shows a student repeatedly inputting the entire code and receiving AI-generated responses with modifications that do not meet the student's needs. The upper half of the figure contains multiple code snippets, while the lower half consists of chatbot responses apologizing for the confusion and attempting to provide solutions. The image represents the student’s frustration as the AI fails to deliver satisfactory results.}
\end{figure*}

Prior experience in specific tasks impacted if and how students choose to use GenAI.
For example, 
in coding tasks, students with more coding experience are more mindful of when to involve GenAI, carefully balancing the relationship between learning outcomes and efficiency when using the tool.\inlinequote{I don't like to just default to it[AI] because I find that I don't think critically or it kinda takes away from the academic experience as a whole [...]. I try to avoid things like that and try to tackle the problem in myself first and only use it for technical, small technical things like libraries and syntaxes that I don't know.}{8}
Less experienced coders focus more on task completion than learning experience. They often lack a clear understanding of the problem when seeking GenAI assistance, leading them to pose vague questions (see~\autoref{fig:prompt-failure}). This often resulted in lower-quality responses, which fall short of expectations, creating a cycle of back-and-forth interactions that generate poor outcomes and negative co-design experiences.\inlinequote{Responses became repetitive and less helpful, particularly with bug fixes in code, often just regurgitating the erroneous code provided.}{10}.

\paragraph{\textbf{Engagement}}
Another critical factor influencing students' decisions to use AI is the level of engagement, shaped by the task's interest and alignment with their learning objectives. When tasks capture students' attention and prompt deeper involvement—such as treating brainstorming as an important and creative process in the course—they tend to complete the work without AI assistance.\inlinequote{ We didn't think about using chat GPT to do that. Like I said, maybe we should do that, but no, it's just talking to each other and trying brainstorming. We just open up a Google Doc and meet in person, and then everyone can type in ideas and we'll discuss it.}{2} In contrast, when tasks are perceived as less relevant to their learning goals, such as documentation, students are more inclined to incorporate AI assistance. \inlinequote{We use it like to fix the grammar of our write-up. Our group was not native speakers. We used it as a tool like Grammarly to fix any grammatical errors in our write-up and also for changing the wordings to make it more fluent.}{4}

\paragraph{\textbf{Process}}
The final significant factor influencing students' sentiments towards GenAI was its comparison against conventional HCI design processes. GenAI, through the most common interaction way of a chatbot, can take away from traditionally not turn-based but slow reflective practices, such as painting. Students often compared GenAI to other tools and methods they used in the design process, such as collaborative whiteboard tools, digital and physical painting tools, forums, web searches, spell checkers,  or manual work. When students found alternative processes to be more efficient, more pleasurable, or more insightful, their sentiment suffered. As one student said: \inlinequote{When I drew, it made me explore deeper in design. Physically as I'm drawing, I'm physically imaging [...] would I actually, want this on my wrist? Why am I drawing it here? If I'm generating something. It's like too many questions to ask at once, perhaps.}{9}.

\subsection{Impact of GenAI on HCI Technology Design and Education - Students' Perspectives}\label{sec:impact}

\subsubsection{\textbf{Students Adherence to the Class Policy}}
The use of Generative AI (GenAI) by students was openly discussed in class; however, it was rarely documented in formal submissions. None of the students sought prior approval from the instructor to utilize GenAI. A thorough review of interview participants' repositories revealed that only one group explicitly documented their use of GenAI for coding, generating visual content (DALL-E), and using for their prototypes' interactivity.
\begin{tcolorbox}[width=\textwidth, colframe=gray, colback=gray!10, title=GenAI Acknowledgement in Group 1's Design Prototype Repository:, sharp corners=south] \footnotesize
 \ttfamily
    Lab 1: Some of our script was coded using the assistance of ChatGPT. \\
    Lab 2: Thank you OpenAI for creating Dall-E, which we used to create the storyboard (+X's Photoshop skills). 
    Chat GPT was used for help in the initial code setup as well as for helping with troubleshooting when we were blocked. \\
    Lab 3: SpotiPi sends the user's response to GPT API. This parses user responses and returns the mood.
    \end{tcolorbox}
The remaining 8 interview groups did not record their use of GenAI in their design repositories, contrary to class policy~(Box: \ref{box:policy}).

When asked about this, students' mentioned 
\inlinequote{We don't necessarily just copy and paste whatever it has generated over. - So it does have some credits, you know. But the thing is you don't you don't give credits to Google.}{6} indicating they did not seek advance permission or disclosed its use because they considered GenAI as a tool auxiliary to their work, similar to a calculator or StackOverflow, rather than a form of cheating.

Students' self-reported engagement with GenAI in the study interviews was characterized by a lack of clarity and consistency. Initially, many students denied using GenAI, citing minimal edits or an inability to recall its application, including whether their teammates had utilized it. However, as they reviewed their class submissions collectively, their memories were prompted, leading to admissions of GenAI usage. Interestingly, when examining code, brainstorming use cases, or final documentation, students struggled to attribute contributions accurately or to determine the extent of GenAI's involvement. This thorough collaborative analysis of class submissions ultimately revealed that students had employed GenAI more extensively than they had disclosed at the interview start. 
\inlinequote{I didn't remember using ChatGPT for lab one. I think we wrote everything here ourselves (Looks at assignment). Oh, this is, -I- didn't generate these images but these were generated by some AI tool. I don't remember. Some group member use Generative AI. [Interviewer: The whole storyboard or parts of the background?] Uh the whole storyboard for storyboard three and four I think.}{10} The lack of definite authorship was seen as critical by students:
\inlinequote{It seems very difficult to sort of untangle. And I think particularly the lack of tools to detect whether, you know, anything was generated by AI or not, right? Whether [students] are doing their own work or not, maybe you deal with that by just making them do it in person in the same way that, you know, math tests are now, you can't sort of trust that students won't just use calculators to get the answers, right?}{1}

\subsubsection{\textbf{Students Overall Performance}}
Students' grades improved, suggesting a better understanding of the material, more efficient workflows, or improved student ability to mislead graders. 
While the change in graders and teaching assistants every year hinders an objective analysis of the grade distribution changes, there is evidence suggesting an improvement in student performance in this class relative to previous years. This is supported by instructors' evaluations of final project prototypes and lab submissions available in public GitHub repositories. Notably, this year's assignments exhibited a significantly higher level of completeness compared to those from prior years, as well as teaching assistants reporting a lighter workload assisting students with prototyping.
In terms of final letter grades, the year with the public release of ChatGPT saw a significant jump in average grade, with 79.17\% of students achieving grades in the A range, showing an increase from 61.22\% the previous year and 62.50\% the year before. 
In summary, student interviews and observations reveal that AI accelerates student work, particularly in the Development phase. Experienced and novice students (considering their coding and design proficiency detailed in \autoref{tab:participants}) alike benefit from the integration of GenAI tools in their workflows. Some groups say outright that GenAI can help them \inlinequote{save time}{1,3,5,6,7}, e.g., how GitHub Copilot automatically assisted them in completing tasks more efficiently. \inlinequote{At that point, I guess we got (...) co-pilot to fill in with [the] long parts. Let's just [do that] right, like because this is so meticulous}{7}. Some students use GenAI to handle tedious tasks.\inlinequote{It helped us understand the math involved because there was a lot of math when it came to rotating the clock}{1}. %
In contrast, some students also report spending more time on tasks, as they had to learn how to use GenAI and then refine its outputs. \inlinequote{The provided answers are less helpful. They just start to provide us with repetitive answers such as this is buggy code [that I] gave to them. It was sometime during the class when they updated the algorithm and it felt like the code quality is getting worse.}{5}, indicating inconsistent usefulness of the tool. Finally, students report what could be interpreted as overreliance on some GenAI tools
\inlinequote{We hit the limit a lot. Yeah. I'll tell you that. Cause we pay for GPT four and there's a three hour limit of like 40 messages that often we hit that on pretty much all of our accounts.-  But I don't want to give GPT all the credit for our work. }{1}. 

\subsubsection{\textbf{Accessibility of GenAI tools}}
Students who paid for the subscription version of ChatGPT reported higher satisfaction and engaged in a broader range of applications compared to those using the free version, who primarily employed it for basic tasks such as spelling and grammar correction. 
\inlinequote{I'm thinking of it because many of my friends who use it really recommend it, [...] it's not minor improvements, but pretty radical improvement.}{4}
Many students noted encountering usage limits with both the free and paid versions. For some, the monthly subscription represented a considerable expense, posing a barrier for students from low-income backgrounds who accessed the university through financial support. \inlinequote{Yeah, so I actually only use Bard and GPT 3.5. Because I'm poor.}{3}

\subsubsection{\textbf{Students' Perception of GenAI tools}}
\paragraph{\textbf{GenAI as a Tool, Collaborator or Teaching Assistant}}

We find that students' perceptions of GenAI roles show significant divergence, with some changing their views on AI across different tasks (I6). Some students tend to personify GenAI, attributing autonomy and emotional connection to it, viewing it as a classmate, a team member, a best friend (I1) or a 24/7 teaching assistant (I6). 
\inlinequote{I mean sh*t, I talk to this thing like it's my best friend sometimes. I ask it a lot of questions. I'll ask it random questions.}{1}. This reflects their perception of AI as a collaborative entity. 
\inlinequote{We would kind of just say - hey let's ask Chatty.}{1}
\inlinequote{It is also like a virtual TA. We might we might code when you guys are offline or asleep. [...] I come from a public university. Sometimes up to a hundred students need to fight for one TA's available time slots.}{6}

\paragraph{\textbf{ChatGPT's Contribution}} In reflecting on the credit given to GenAI tools for class work, students offered varied perspectives. In the interview, students were asked "If you had \$1000 to distribute across your class groupmates and ChatGPT, how much would you allocate to ChatGPT?" (See also \autoref{tab:participants}). Some groups (I1, I3, I6, I9) saw AI as a full-fledged team member and chose to divide the bonus equally\inlinequote{Yeah, it would be equal. If we're 5 members and one of them is GPT, it probably get a \$200 share.}{3}.
In contrast, other students described it as "a complementary tool" (I5), "a cool tool" (I8) or "a better tool"(I10) and downplay it's role as a mere instrument, similar to a calculator(I2, I3), using Google (I3, I6) or StackOverflow (I4, I5, I6).
From the hypothetical price points a few interview groups gave GenAI a low share (I2, I4, I5, I6, I7) which was in some cases still higher than some of the human team member attributions, as one student stated: 
\emph{``I would give \$950 to myself and \$50 to [my team member].''}, and attribute GenAI a sum that was seven times what the human teammate received, illustrating the disparity between their valuation of GenAI and certain human contributors.

\subsubsection{\textbf{Learning and Ethical Reflection}}\label{ssec:Ethical Concerns}
Students repeatedly raised concerns about the ethical implications of using GenAI in their design work in the context of learning and design reflection. They expressed apprehension about GenAI hindering their learning and problem-solving skills, as well as its potential to deceive graders and professors. Students were worried that over-reliance on GenAI might undermine their understanding of human-centered design and the design process itself, leading to misuse in learning contexts, as the following quotes illustrate (shortened for brevity):

\begin{quote}
\textit{P1:} "It's a tool, and people are either going to use it to learn or just get by. And I've done both."  

\textit{P3:} "It could take away the thinking process."  

\textit{P1:} "Yeah, I've seen people just have it think for them. Type in the question, get the answer, copy-paste, submit."  

\textit{P2:} "That's nasty."  

\textit{P1:} "Yeah, I don't agree with that."  

\textit{Interviewer:} "How do you avoid it? Do you want to avoid it?"  
\textit{P1:} "For me, it's a mindset. I'm here to learn for the most part. I guess it's just a matter of motivation for people, at least for me it is."  

\textit{P3:} "Priorities.  You're paying so much to learn. You shouldn't be taking away the thinking process that you paid so much money for. People say you go to college to change your way of thinking. " 

(I1: P1, P2, P3)
\end{quote}

\section{Discussion}

In this study, we provided an overview of students' self reported use of GenAI tools for an HCI design class, showcasing the broad set of use cases, usage patterns, and sentiment towards these tools. 
While students acknowledged the utility of GenAI to make task completion easier and supporting their academic journey, this practice does not necessarily enhance students' comprehension. Among the four ways students utilize GenAI (see \autoref{sssec:usage-taxonomy}), both the Amplifier and Executor usage patterns tend to produce outcomes beyond students' own capabilities, bypassing the learning and critical reflection that typically comes from effort, which could significantly impact their learning outcomes. In executor pattern cases, delegation is often viewed positively and can free up time for other tasks; however, in many cases, GenAI hinders depth in design. For example, a key learning objective in design classes is to avoid idea fixation~\cite{kohnCollaborativeFixationEffects2011}, often enforced through exercises such as generating 100 ideas, including bad ones, to increase the likelihood of discovering a few truly innovative concepts~\cite{innovationtraining100IdeasExercise2023}. 
GenAI drastically reduces the cost and effort associated with idea generation, potentially depriving students of the deeper and broader benefits of brainstorming. Critically, GenAI challenges such heuristics for evaluating subjective quality aspects that provide graders with indications of effort through quantity.

The emergent findings presented here have implications on how we should think and plan HCI design education and tools.%

\subsection{Prompt Need To Adapt HCI Education for the Era of GenAI}
\begin{figure}[h]
    \centering
    \includegraphics[width=0.5\textwidth]{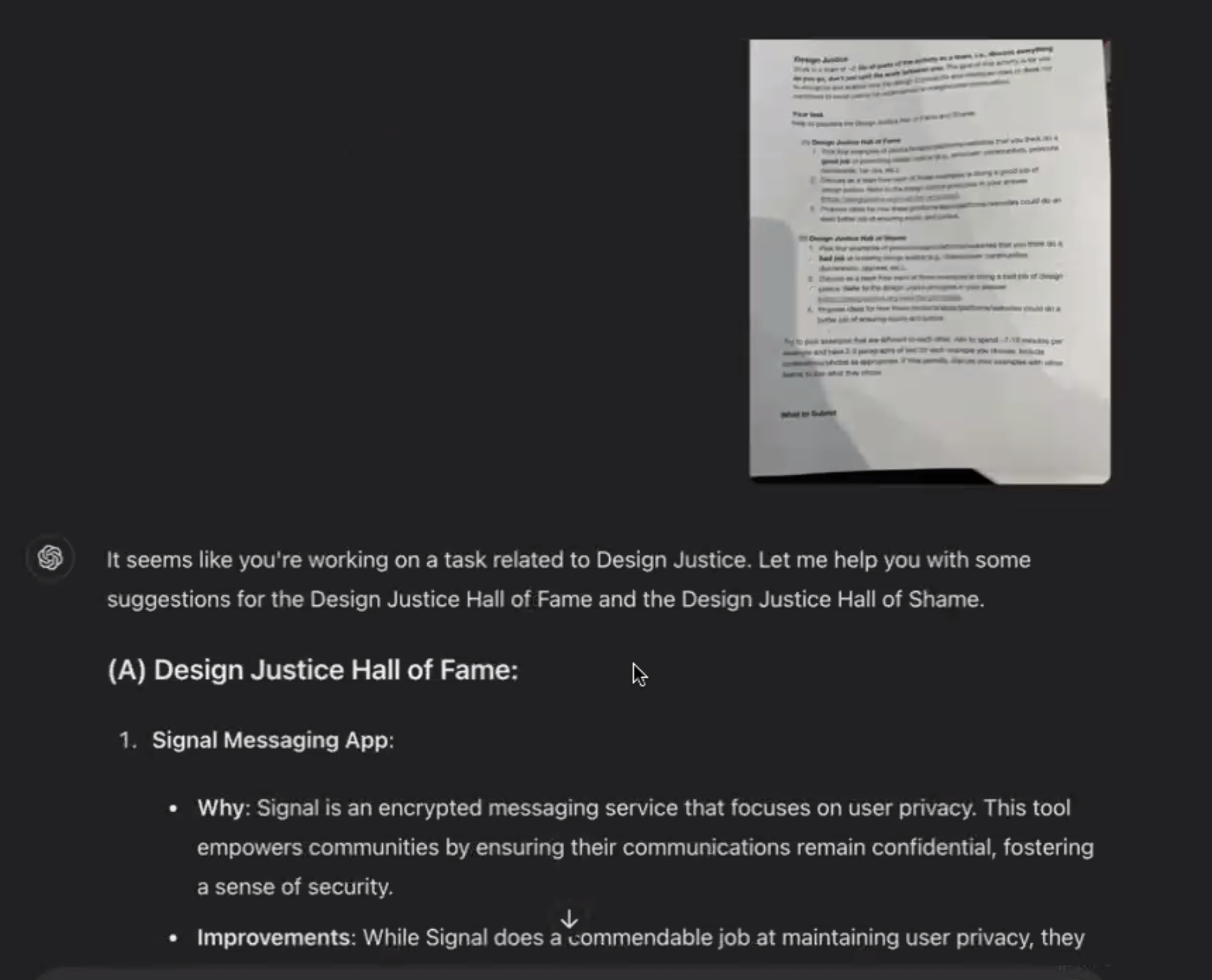}
    \caption{Some courses attempt to mitigate the impact of GenAI by implementing paper-based assessments, which may inadvertently encourage academic dishonesty. The image depicts a vision-enabled Generative AI conversational tool completing a homework assignment shared by a student from a different HCI course, where participants expressed \emph{curiosity} about the AI's ability to fulfill assignment requirements.}
    \label{fig:cheating}
    \Description{The figure illustrates an example of a homework assignment being completed by a vision-enabled Generative AI. The assignment, shared by a student from a different HCI course, is shown as a piece of paper with text, which the AI is scanning and solving. This represents concerns raised by students in interviews about the potential for Generative AI to complete academic work, including assignments that would traditionally require human effort and understanding.}
\end{figure}

The interview study shows that GenAI is going to be an integral part of novice HCI practitioners process, thus HCI education needs to adapt to it. 
While related work has discussed educational concerns and educators initial response was to "ban it 'till we understand it"~\cite{lau_ban_2023}, students shared anecdotal evidence of the length they go to use these tools, with students reporting about GenAIs ability to fully answer printed HCI Design assignments (see~\autoref{fig:cheating}). While in this interactive device design class, GenAI use appeared mostly beneficial; interviewees indicated that the use of GenAI in other HCI classes focusing on deep understanding of material and less so generated output, poses more significant challenges, with generated questions often lacking specificity and failing to elicit deeper insights. 
This raises concerns about potential oversimplification, which could lead to a superficial understanding of the subject matter. As one participant noted,"\inlinequote{Now you're learning less in HCI class. ... To code up an interview script, if you don't have ChatGPT you need to read over the script.  ... Everyone in the class, they just dump the script into ChatGPT and try to generate a result.}{5}

As the GenAI tools are becoming more powerful, they are likely to become useful in more contexts. The qualitative difference reported by students between model versions, and the continued use of GenAI for the specific use cases indicates so. As such, it is important that HCI education adapts to this new reality. The results of this study show that while the use of GenAI tools in the classroom has mixed efficiency, HCI courses can benefit and should rethink learning outcomes to ensure GenAI tools have a positive impact on student's learning outcomes and use of the technology where co-design becomes the status quo.
\inlinequote{At the least and at most, education has changed based on AI, and allowing it in some way is necessary because straight up banning it is just impossible... Everything has to be reviewed and possibly reimagined to fit what we want students to focus their time most on}{3}

\subsubsection{\textbf{Curriculum Adaptation}}

This study evidences that HCI educators have to modify their curricula to incorporate GenAI tools, as students expressed a commitment to continue using them in their design work. This will have to involve teaching students how to use GenAI tools effectively and discussing their ethical implications. Educators should also consider how to grade students work when employed to complete assignments. One approach is to design tasks that require students report the use  of both GenAI and their own thinking, rather than simply generating answers. 
Here the the taxonomy of GenAI usage patterns is helpful. As a \emph{Booster}: require students to document original input and how GenAI was used to refine it. As a \emph{Benchmark}: demand students to include GenAI output (for example in brainstorming) and document how students diverged beyond that. 
Often, the option needs to be to demand more: when ideas can be generated so cheaply, asking for a handful of ideas is not sufficient. Instead, ask students to generate some with AI, some by themselves, and some through research. Encourage students to not only list the items but also to critically reflect on them, remix and synthesize their ideas. Additionally, many pen and paper design thinking methods allow for side-stepping GenAI. We recommend relying more on tangible divergent design thinking methods, such as bodystorming, role play, or card-based brainstorming~\cite{Hornecker2010-tq}.

One of the most promising aspects of GenAI integration is the potential for improved iteration speed, which could allow students to spend more time on testing and refining their designs rather than just implementing basic functionality. By accelerating certain aspects of the design process, students might engage in more iterations and gather more user feedback, potentially leading to better-designed outcomes. However, this requires deliberate curriculum design that emphasizes the importance of these later-stage activities and ensures that time saved in implementation is redirected toward deeper engagement with users and reflection on design choices.

\subsubsection{\textbf{Class Policy}}
While the adopted class policy failed to get students to appropriately disclose or document their use of GenAI tools, it is important to balance students' fear of lower grades with the need for transparency.
Students did not think about GenAI doing the work for them, but rather auxiliary tools. Some educators may choose to ban the use of these tools, while others may allow them with certain restrictions. We recommend educators to clearly communicate their policies on the use of GenAI tools throughout the class, as for each individual assignment, different uses can be appropriate. 
They should also consider how to incorporate grading with students' use of these tools and ensure that they can be adequately used to complete assignments or exams, without pushing students towards dishonesty. Educators should promote a climate of trust to learn about appropriate co-design with GenAI rather than threatening repercussions. 

\subsubsection{\textbf{Financial Accessibility}}
\setcounter{footnote}{0}
\renewcommand*{\thefootnote}{\fnsymbol{footnote}}
Current research in HCI and education rarely explores the financial impact of GenAI tools. As these tools become more prevalent in education but costly subscriptions are released (with ChatGPT pro being \$200 per month\footnote{at time of writing}), it is essential to establish clearer policies on which versions should be used. To ensure more equitable access, educational institutions could consider subsidizing paid versions or incorporating GenAI tools into their resources. Doing so would promote academic fairness and prevent financial barriers from limiting students' ability to engage with advanced technologies. One way to manage GenAI usage in HCI design and prototyping classes may be to provide access to a free class-specific GenAI account or API. This would ease access and equity barriers, allow students to use the tool in a controlled environment, and would enable educators to monitor students' use of the tool. Further, such a class-specific GenAI tool could enforce or remind students about the class policy. This could help prevent academic dishonesty and ensure that students are using the tool effectively. Based on this class's student reflections, it could incorporate reminders to use GenAI for setting benchmarks to compare creativity to, rather than as a final solution. Implementing a class-specific GenAI chatbot for Human-Computer Interaction design and development can also reduce student expenses, as credit-based subscriptions are considerably more economical for a class than for individual students. Furthermore, emerging models such as GPT-4o mini are virtually cost-free for many tasks undertaken by students~\cite{franzenOpenAIUnveilsGPT4o2024}. %
\renewcommand*{\thefootnote}{\arabic{footnote}}

\subsection{\textbf{Customization of GenAI Tools}}

The results of this study suggest that GenAI tools could be customized to better meet the needs of interaction design students. This could involve developing tools that are specifically designed for use in HCI courses, with features that support students' learning and help them develop their skills beyond automating tasks. For example, tools could be designed to generate questions that are more specific and relevant to the subject matter or to provide feedback on students' use of the tool. For this class, a core limitation of GenAI, according to students, was its lack of knowledge about hardware topics and pre-defined libraries by the assignment's prototyping scripts. A potential solution are fine-tuned chatbots, or chatbots with access to a vector database of specific class libraries and knowledge about available class hardware components. %
For example, a main barrier to students' use of contextual code assistance was the use of coding on Raspberry PIs rather than on students' own hardware. This made use of integrated GenAI coding support tools such as Github Copilot more laborious, which can be seen both as a challenge (barrier to full automation of prototyping) or as a learning opportunity (develop a GenAI-supported coding and hardware prototyping environment), based on intended learning objectives. Finally, beyond chatbots for coding and hardware support, GenAI can be customized for the whole HCI technology design process in education. As we now know about the limitations of current out-of-box GenAI for brainstorming, design reflection, and learning support, we are better informed to imagine \emph{purposeful, and appropriate} GenAI use.  %
Interestingly, our findings indicate that the probabilistic nature of GenAI~\cite{yang_re-examining_2020} may be less crucial to student experiences than their approach to using GenAI and their prompting strategies ~\cite{Zamfirescu-Pereira2023-sj}. Rather than waiting for new types of AI systems specifically designed for education and critical thinking, we suggest adapting current GenAI tools through thoughtful prompt guides and usage patterns that encourage deeper engagement with design principles. Our usage taxonomy could inform the development of prompting strategies that emphasize benchmarking and boosting rather than executing and amplifying, thereby supporting rather than supplanting critical design thinking.

\subsection{Limitations}
While students' grades improved notably compared to the previous years, this increase could also be influenced by factors such as changes in teaching assistants and graders.

The study's findings are based on interviews with master's students from a single class comprising individuals with diverse academic backgrounds, which may limit their generalizability to other student populations or HCI courses. However, the authors are reasonably confident that the experiences reported here are applicable to similar classes, even with variations in software and hardware (e.g., using Arduino instead of Raspberry Pi) and the context of industry technology design. Additionally, the insights are considered to reflect a broader and enduring shift, as large language models based on the transformer architecture approach a plateau in their development. Educators across institutions will need to familiarize themselves with these tools and consider which learning outcomes they want to prioritize and which tasks can be safely delegated to GenAI. Student sentiments may evolve over time, but our taxonomy is designed to remain robust and adaptable for future co-design and co-prototyping efforts. 

\subsection{Future Work}
In the context of the prototyping and design class examined, the benefits of the GenAI tool for students are evident; however, this may not be the case in other HCI courses. We posit that research-intensive classes may not derive the same advantages from the tool as design and prototyping courses and, in some instances, the drawbacks may overshadow the benefits. While there is already a growing corpus of literature on these topics, future research should further explore the application of GenAI tools in various HCI contexts, particularly in understudied areas of UX research, UX writing, and throughout the HCI design cycle, to assess whether the tool's advantages are concordant across these settings.

The prototyping design class studied here appears relatively prepared for the arrival of GenAI, but a class that specifically designs its content for the arrival and potential of GenAI tools could be even more beneficial. This could involve a more in-depth exploration of the tools' capabilities, limitations, and ethical considerations, as well as the development of a curriculum that integrates GenAI into the learning process. %
Future research should investigate the effects of such a curriculum and adapted tools on students' learning outcomes and their proficiency in applying Generative AI tools in practice.

\section{Conclusion}

GenAI tools already enhance efficiency and can foster creativity in Interactive System Design with a strong potential to hamper students' learning experiences and critical reflection skills. Balancing GenAI use with deliberate, nuanced engagement in co-design is crucial. This study unpacks the usage patterns and factors influencing GenAI's success or failure when used by novice technology designers. Adjusting HCI classes to account for GenAI's impact is essential, with a focus on amplifying its empowering aspects while addressing and mitigating the aforementioned drawbacks.

Looking to the future, HCI education must prepare students for a landscape where GenAI co-design becomes standard. With the likelihood of task automation in industry, it's essential to emphasize a practice-oriented approach that nurtures skills beyond GenAI’s reach, such as critical evaluation, ethical judgment, and deep understanding of human needs.

\begin{acks}
We thank all students for sharing their experience using GenAI in the interactive device design 2023 fall class at Cornell Tech. We thank Prof. Shiri Azenkot and Madiha Zahrah Choksi for their insights and discussions about students' use of GenAI in UX writing and UX research. 
\end{acks}

\bibliographystyle{ACM-Reference-Format}

\bibliography{GenAI-prototyping-class,manual-references}

\newpage
\appendix
\label{sec:appendix}
\section{Students Sentiment About Each GenAI Use Case per Interview}

\begin{figure}[h]
\centering
    \includegraphics[width=0.95\linewidth]{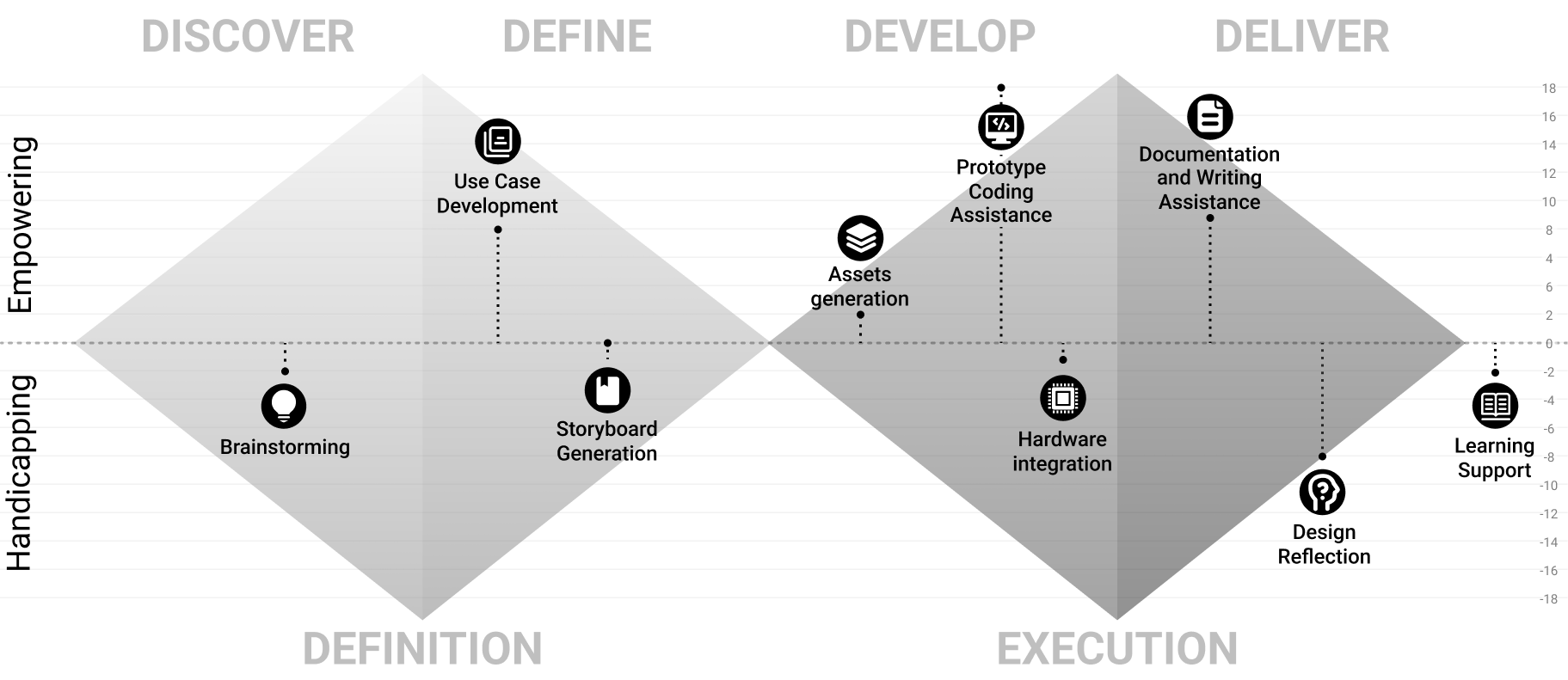}
    \caption{Students' use of generative artificial intelligence mapped to the Double Diamond technology design process~\cite{designcouncilDoubleDiamondDesign2004}. Students' ordinal sentiment about these use cases is quantified from \autoref{tab:sentiment-use-cases} codes.}
    \label{fig:double-diamond}
    \Description{This diagram maps the student's use of generative artificial intelligence across the Double Diamond design process model, which consists of four phases: Discover, Define, Develop, and Deliver. The model is visualized with two diamonds, each divided into two parts. The left diamond, labeled 'Definition,' includes phases Discover and Define, focusing on tasks like brainstorming, use case development, and storyboard generation. The right diamond, labeled 'Execution,' covers the Develop and Deliver phases, detailing tasks like asset generation, prototyping assistance, hardware integration, documentation assistance, and design reflection. Another use case depicted outside of the diamond is learning support.}
\end{figure}

\renewcommand{\arraystretch}{1.5} %

\begin{table}[hh] 
\resizebox{\textwidth}{!}{%
\begin{tabular}{@{}r|>{\raggedright\arraybackslash}p{1.85cm}|>{\raggedright\arraybackslash}p{1.85cm}>{\raggedright\arraybackslash}p{1.85cm}|>{\raggedright\arraybackslash}p{1.85cm}>{\raggedright\arraybackslash}p{1.85cm}>{\raggedright\arraybackslash}p{1.85cm}|>{\raggedright\arraybackslash}p{1.85cm}>{\raggedright\arraybackslash}p{1.85cm}|>{\raggedright\arraybackslash}p{1.85cm}@{}}
\toprule
\multicolumn{1}{r|}{Design Phase} &
  \cellcolor[HTML]{FFFFFF}Discover &
  \cellcolor[HTML]{FFFFFF}Define &
  \cellcolor[HTML]{FFFFFF} &
  \cellcolor[HTML]{FFFFFF}Develop &
  \cellcolor[HTML]{FFFFFF} &
  \cellcolor[HTML]{FFFFFF} &
  \cellcolor[HTML]{FFFFFF}Deliver &
  \cellcolor[HTML]{FFFFFF} &
  \cellcolor[HTML]{FFFFFF}Learning \\ \midrule
\multicolumn{1}{l|}{\makecell[r]{\vspace{0.45cm}\textbf{Use Case /}\\ \textbf{Interview:}}} &
  \cellcolor[HTML]{FFFFFF}\textbf{Brain-storming} &
  \cellcolor[HTML]{FFFFFF}\textbf{Use case development} &
  \cellcolor[HTML]{FFFFFF}\textbf{Storyboard generation} &
  \cellcolor[HTML]{FFFFFF}\textbf{Assets generation} &
  \cellcolor[HTML]{FFFFFF}\textbf{Prototype coding assistance} &
  \cellcolor[HTML]{FFFFFF}\textbf{Hardware integration assistance} &
  \cellcolor[HTML]{FFFFFF}\textbf{Docu-mentation assistance} &
  \cellcolor[HTML]{FFFFFF}\textbf{Design reflection} &
  \cellcolor[HTML]{FFFFFF}\textbf{Learning support} \\ \midrule 
1 &
  \cellcolor{Underperforming}Under-performing &
  \cellcolor{NotMentioned} &
  \cellcolor{AssistedSuccess}Assisted Success &
  \cellcolor{AssistedSuccess}Assisted Success &  \cellcolor{AssistedSuccess}Assisted Success &
  \cellcolor{UncertainPerformance}Uncertain Performance &
  \cellcolor{NotMentioned} &
  \cellcolor{QuestionableIntegrity}Ethical Concerns &
  \cellcolor{UncertainPerformance}Uncertain Performance \\ 
2 &
  \cellcolor{NotMentioned} &
  \cellcolor{NotMentioned} &
  \cellcolor{Underperforming}Under-performing &
  \cellcolor{AssistedSuccess}Assisted Success &
  \cellcolor{NotMentioned} &
  \cellcolor{NotMentioned} &
  \cellcolor{NotMentioned} &
  \cellcolor{NotMentioned} &
  \cellcolor{QuestionableIntegrity}Ethical Concerns \\
3 &
  \cellcolor{AssistedSuccess}Assisted Success &
  \cellcolor{NotMentioned} &
  \cellcolor{Underperforming}Under-performing &
  \cellcolor{Underperforming}Under-performing &
  \cellcolor{AssistedSuccess}Assisted Success &
  \cellcolor{Underperforming}Under-performing &
  \cellcolor{AssistedSuccess}Assisted Success &
  \cellcolor{NotMentioned} &
 \cellcolor{QuestionableIntegrity}Ethical Concerns \\
4 &
  \cellcolor{NegativeUtility}Negative Utility &
  \cellcolor{AssistedSuccess}Assisted Success &
  \cellcolor{NotMentioned} &
  \cellcolor{NotMentioned} &
  \cellcolor{AssistedSuccess}Assisted Success &
  \cellcolor{NotMentioned} &
  \cellcolor{AssistedSuccess}Assisted Success &
  \cellcolor{NotMentioned} &
  \cellcolor{NotMentioned} \\
5 &
  \cellcolor{NegativeUtility}Negative Utility &
  \cellcolor{NotMentioned} &
  \cellcolor{AssistedSuccess}Assisted Success &
  \cellcolor{NegativeUtility}Negative Utility &
  \cellcolor{UncertainPerformance}Uncertain Performance &
  \cellcolor{NegativeUtility}Negative Utility &
  \cellcolor{HighEfficacy}High Efficacy &
  \cellcolor{AssistedSuccess}Assisted Success &
 \cellcolor{QuestionableIntegrity}Ethical Concerns \\
6\textsuperscript{*} &
  \cellcolor{AssistedSuccess}Assisted Success &
  \cellcolor{AssistedSuccess}Assisted Success &
  \cellcolor{NotMentioned} &
  \cellcolor{NotMentioned} &
  \cellcolor{AssistedSuccess}Assisted Success &
  \cellcolor{Underperforming}Under-performing &
  \cellcolor{Underperforming}Under-performing &
  \cellcolor{AssistedSuccess}Assisted Success &
  \cellcolor{HighEfficacy}High Efficacy \\
7\textsuperscript{*} &
  \cellcolor{NotMentioned} &
  \cellcolor{NotMentioned} &
  \cellcolor{Underperforming}Under-performing &
  \cellcolor{NotMentioned} &
  \cellcolor{AssistedSuccess}Assisted Success &
  \cellcolor{Underperforming}Under-performing &
  \cellcolor{AssistedSuccess}Assisted Success &
  \cellcolor{NotMentioned} &
  \cellcolor{NotMentioned} \\
8 &
  \cellcolor{Underperforming}Under-performing &
  \cellcolor{NotMentioned} &
  \cellcolor{NotMentioned} &
  \cellcolor{NotMentioned} &
  \cellcolor{AssistedSuccess}Assisted Success &
  \cellcolor{Underperforming}Under-performing &
  \cellcolor{NotMentioned} &
  \cellcolor{NotMentioned} &
  \cellcolor{UncertainPerformance}Uncertain Performance \\
9\textsuperscript{†} &
  \cellcolor{AssistedSuccess}Assisted Success &
  \cellcolor{AssistedSuccess}Assisted Success &
  \cellcolor{NegativeUtility}Negative Utility &
  \cellcolor{Underperforming}Under-performing &
  \cellcolor{UncertainPerformance}Uncertain Performance &
  \cellcolor{AssistedSuccess}Assisted Success &
  \cellcolor{NotMentioned} &
  \cellcolor{QuestionableIntegrity}Ethical Concerns &
  \cellcolor{NotMentioned} \\
10 &
  \cellcolor{Underperforming}Under-performing &
  \cellcolor{AssistedSuccess}Assisted Success &
  \cellcolor{AssistedSuccess}Assisted Success &
  \cellcolor{AssistedSuccess}Assisted Success &
  \cellcolor{UncertainPerformance}Uncertain Performance &
  \cellcolor{Underperforming}Under-performing &
  \cellcolor{NotMentioned} &
  \cellcolor{QuestionableIntegrity}Ethical Concerns &
  \cellcolor{HighEfficacy}High Efficacy \\
11 &
  \cellcolor{NotMentioned} &
  \cellcolor{NotMentioned} &
  \cellcolor{NotMentioned} &
  \cellcolor{NotMentioned} &
  \cellcolor{AssistedSuccess}Assisted Success &
  \cellcolor{AssistedSuccess}Assisted Success &
  \cellcolor{NotMentioned} &
  \cellcolor{NotMentioned} &
  \cellcolor{AssistedSuccess}Assisted Success \\
12\textsuperscript{†} &
  \cellcolor{Underperforming}Under-performing &
  \cellcolor{NotMentioned} &
  \cellcolor{Underperforming}Under-performing &
  \cellcolor{NotMentioned} &
  \cellcolor{UncertainPerformance}Uncertain Performance &
  \cellcolor{UncertainPerformance}Uncertain Performance &
  \cellcolor{UncertainPerformance}Uncertain Performance &
  \cellcolor{QuestionableIntegrity}Ethical Concerns &
  \cellcolor{QuestionableIntegrity}Ethical Concerns \\
 \midrule

\end{tabular}%

\begin{minipage}[t]{0.35\textwidth}
\small
\raggedleft
\begin{tabular}{>{\raggedright\arraybackslash}m{0.89\textwidth}}
\textbf{Legend:} \\ \vspace{-5pt} The categories are determined by the following codes. The table containing transcript quotes is accessible on OSF. \\
\textbf{\fcolorbox{HighEfficacy}{HighEfficacy}{  High Efficacy}}: GenAI demonstrates high efficacy in task execution, meeting student expectations with satisfactory results.\\
\textbf{\fcolorbox{AssistedSuccess}{AssistedSuccess}{  Assisted Success}}: GenAI is beneficial with human intervention, serving as a valuable foundation or supplementary tool.\\
\textbf{\fcolorbox{UncertainPerformance}{UncertainPerformance}{  Uncertain Performance}}: GenAI exhibits mixed performance, providing useful assistance in some areas while falling short in others.\\
\textbf{\fcolorbox{Underperforming}{Underperforming}{  Under-performing}}: GenAI exhibits average performance, often failing to meet student expectations.\\
\textbf{\fcolorbox{NegativeUtility}{NegativeUtility}{Negative Utility}}: GenAI introduces additional complexity and issues during use, with students expressing dissatisfaction with its outputs.\\
\textbf{\fcolorbox{QuestionableIntegrity}{QuestionableIntegrity}{  Ethical Concerns}} Although GenAI can produce work of seemingly good quality, students raise concerns that this use could potentially be misleading, including to graders.\\
\textbf{\fcolorbox{NotMentioned}{NotMentioned}{  Not Mentioned}}: There is no mention of this use case using ChatGPT or other GenAI tools.
\end{tabular}
\end{minipage}
}
\caption{Interviewees' sentiments on using Generative AI for HCI use cases. Asterix\textsuperscript{*} and dagger\textsuperscript{†} indicate interviews with students from the same final groups. The spreadsheet containing interview transcript excerpts for assessing student sentiment is accessible in the Open Science Foundation repository~\cite{Sandhaus_Gu_Parreira_Ju_2024}.}
\label{tab:sentiment-use-cases}
\Description{
This table shows the interviewees' sentiment regarding their use of Generative AI (GenAI) for various stages of HCI tasks across 12 interviews. The columns represent design phases (Discover, Define, Develop, Deliver, and Learning), and different GenAI use cases such as brainstorming, use case development, storyboard generation, asset generation, prototype coding assistance, hardware integration assistance, documentation assistance, design reflection, and learning support. The rows correspond to each interview, and cells are color-coded based on the sentiment categories assigned: High Efficacy, Assisted Success, Uncertain Performance, Underperforming, Negative Utility, and Ethical Concerns. Each cell indicates how the interviewees felt about the use of GenAI for that particular task during the design phase. The table includes a legend explaining each sentiment. Ethical concerns are marked for tasks where students expressed worries about the potential consequences of using GenAI. The table notes that a spreadsheet with detailed transcripts is available on the Open Science Foundation repository.
}
\end{table}

\end{document}